\documentclass[11pt]{article}

\usepackage{geometry}  % Flexible and complete interface to document dimensions
\usepackage[T1]{fontenc}  % Standard package for selecting font encodings
\usepackage[utf8x]{inputenc}  % Accept different input encodings
\usepackage{lineno}  % Line numbers on paragraphs
\usepackage{fancyhdr}  % Extensive control of page headers and footers in LaTeX2ε
\usepackage{enumitem}  % Control layout of itemize, enumerate, description
\usepackage{hyperref}  % Extensive support for hypertext in LaTeX
\usepackage{titling}  % Control over the typesetting of the \maketitle command
\usepackage{natbib}  % Flexible bibliography support
\usepackage{mathtools}  % Mathematical tools to use with amsmath
\usepackage{titlesec}  % Select alternative section titles
\usepackage{lastpage}  % Reference last page for Page N of M type footers

\usepackage[english]{babel}  % Multilingual support for Plain TeX or LaTeX
\usepackage{amsmath}  % AMS mathematical facilities for LaTeX
\usepackage{amsfonts}  % TeX fonts from the American Mathematical Society
\usepackage{amssymb}  % Additional symbols from American Mathematical Society
\usepackage{wasysym}  % LaTeX support file to use the WASY2 fonts
\usepackage{bbm}  % "Blackboard-style" cm fonts
\usepackage{array}  % Extending the array and tabular environments
\usepackage{xr}  % References to other LaTeX documents
\usepackage{verbatim} % Reimplementation of and extensions to LaTeX verbatim
\usepackage{float} % Improved interface for floating objects

\usepackage[dvipsnames]{xcolor}
\definecolor{orcidlogocol}{HTML}{A6CE39}
\newcommand{\orc}{\includegraphics[height=\fontcharht\font`A]{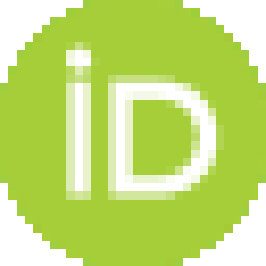}}

\usepackage[skip=1.5pt,font=footnotesize]{caption}

\usepackage{accsupp}% http://ctan.org/pkg/accsupp
\usepackage{lipsum}% http://ctan.org/pkg/lipsum

\usepackage{booktabs}
\usepackage{siunitx}
\usepackage[table]{xcolor}
\usepackage{pdfpages}

\hypersetup{%
  pdfborderstyle={/S/U/W 1}% border style will be underline of width 1pt
}

\setcitestyle{round,numbers}
\titleformat{name=\section}{\normalfont\Large\bfseries}{}{0pt}{}
\titleformat{name=\subsection}{\normalfont\large\bfseries}{}{0pt}{}
\titleformat{name=\subsubsection}{\normalfont\normalsize\bfseries}{}{0pt}{}

\newcommand{\email}[1]{\href{mailto:{#1}}{{#1}}}

\newcommand{\keywords}[1]{\textbf{Keywords}: {#1}}

\newcommand{\optincludegraphics}[2][]{}
\newcommand{\optinput}[1]{}
\newcommand{\capt}[2][]{\caption[{#1}]{\textbf{{#1}}\newline{#2}}}

\newtagform{brackets}{[}{]}
\usetagform{brackets}

\newcommand{\thejournal}[1]{Magnetic Resonance in Medicine}

\title{Controlling spatial correlation in k-space interpolation networks for MRI reconstruction: denoising versus apparent blurring}
\begin{document}

% ======================================================================
%TC:ignore
\begin{titlepage}
{\noindent\LARGE\bf \thetitle}

\medskip%\bigskip

% : Insert author names, affiliations and corresponding author email
% : (do not include titles, positions, or degrees).

\begin{flushleft}\large
	Istvan Homolya\textsuperscript{1,2,3,{*}\hspace{0.5mm}\href{https://orcid.org/0000-0002-0662-6464}{\orc}},
	Jannik Stebani\textsuperscript{2\hspace{0.5mm}\href{https://orcid.org/0009-0004-9631-9928}{\orc}},
	Felix Breuer\textsuperscript{4\hspace{0.5mm}\href{https://orcid.org/0000-0002-9676-9038}{\orc}},
        Grit Hein\textsuperscript{5\hspace{0.5mm}\href{https://orcid.org/0000-0001-5696-6486}{\orc}},
        Matthias Gamer\textsuperscript{6\hspace{0.5mm}\href{https://orcid.org/0000-0002-9676-9038}{\orc}},
        Florian Knoll\textsuperscript{3,7\hspace{0.5mm}\href{https://orcid.org/0000-0001-5357-8656}{\orc}},
        and Martin Blaimer\textsuperscript{4\hspace{0.5mm}\href{https://orcid.org/0000-0002-6360-9871}{\orc}}
\end{flushleft}

\medskip%\bigskip

\noindent

{\small
\begin{enumerate}[label=\textbf{\arabic*}]
\item Molecular and Cellular Imaging, Comprehensive Heart Failure Center, University Hospital Würzburg, Würzburg, Germany
\item Department of Physics, University of Würzburg, Würzburg, Germany
\item Department Artificial Intelligence in Biomedical Engineering, Friedrich-Alexander-Universität Erlangen-Nürnberg, Erlangen, Germany
\item Magnetic Resonance and X-ray Imaging Department, Fraunhofer Institute for Integrated Circuits IIS, Division Development Center X-Ray Technology, Würzburg, Germany
\item Department of Psychiatry, Psychosomatics, and Psychotherapy, University Hospital Würzburg, Würzburg, Germany
\item Department of Psychology, University of Würzburg, Würzburg, Germany
\item Department of Radiology, New York University Grossman School of Medicine, New York, New York, USA
\end{enumerate}
}

\medskip%\bigskip

% : Use the dagger symbol to denote a single equal contribution authorship.
% : Multiple equal-contribution authorship may be included in the acknowledgments.

%\textbf{{†}}: These authors contributed equally to this work.

% : Use the asterisk to denote corresponding authorship.
% : Provide email address in note below.
{\small
\textbf{*} Corresponding author:

\indent\indent
\begin{tabular}{>{\bfseries}rl}
Name		& Istvan Homolya								\\
Department	& Molecular and Cellular Imaging			    \\
Institute	& Comprehensive Heart Failure Center			\\
Address 	& Am Schwarzenberg 15, 97078 Würzburg, Germany	 \\		
E-mail		& \email{istvan.homolya@uni-wuerzburg.de}		 \\
\end{tabular}
}

\medskip
\begin{center}
%\textbf{\textsc{\thejournal}}
\textbf{\textsc{Submitted to \thejournal}}
\end{center}

\vfill

% ======================================================================
% : set word count results (+++ must be included, 1200  must be excluded)
% 	+++ introduction, theory, methods, results, discussion, conclusion,
%		appendix, 
% 	1200  title page, abstract, figure captions, tables, table captions,
%		references, revision markings
% : first argument is the manuscript word count
% : second argument is the abstract word count
% : to use `texcount` results, use '%TC:ignore'/'%TC:endignore' directives.
% : \wcManuscript and \wcAbstract should perform the correct word count.
%\wordcount{\wcManuscript}{\wcAbstract}
%\newline\bigskip\textbf{Number of Figures+Tables:}

% : display detailed word count
%\wcTotal

\begin{tabular}{>{\bfseries}rl}
\textbf{Manuscript word count}: &5073\\
\textbf{Abstract word count}: &250\\
\textbf{Number of Figures $+$ Tables}: &10\\
\end{tabular}

\end{titlepage}
%TC:endignore
% ======================================================================

% ======================================================================
% ======================================================================
\pagebreak
% ======================================================================
% ======================================================================

% ======================================================================
%TC:break Abstract
\begin{abstract}
\textbf{Purpose:} Interpretability is essential for the clinical adoption of state-of-the-art machine learning (ML) methods in magnetic resonance imaging (MRI). Conventional evaluation of ML reconstructions relies heavily on aggregate image metrics that require fully sampled references. These metrics, inherited from classical image processing and natural image ML, often overlook the critical challenge of noise amplification specific to medical image reconstruction. This study aims to analyze the influence of nonlinear activations on spatial noise variance distribution of k-space interpolation networks (RAKI) and to provide a framework for incorporating variance maps during network training.

\textbf{Methods:} We present an analytical framework that decomposes pixel-level noise variance into components reflecting linear and nonlinear characteristics of RAKI. By applying automatic differentiation on the image-space equivalent of the network, variance maps are computed during each training iteration, enabling runtime quality assessment beyond data consistency. We introduce apparent blurring, quantifying nonlinear signal mixing without dependence on reference images. By incorporating variance maps into the traning loss as regularizers, our self-informed RAKI architecture (G-factor-informed RAKI, GIF-RAKI) can directly integrate updated noise characteristics during runtime.

\textbf{Results:} Experimental results demonstrate that variance components quantitatively explain network behavior. GIF-RAKI outperforms conventional RAKI variants in image fidelity and noise suppression.

\textbf{Conclusion:} Our methodology advances practical and theoretical aspects of ML-based MRI reconstruction by reinstating reconstruction noise characterization as a cornerstone for performance evaluation, eliminating the need for fully sampled references. GIF-RAKI also enables optimization of the trade-off between denoising and apparent blurring.
\end{abstract}

% ======================================================================
% : set search-engine keywords (3 to 6)
\bigskip

% ======================================================================
% ======================================================================
\pagebreak
% ======================================================================
% ======================================================================

%TC:break _main_
% ======================================================================
\section{Introduction}
% ======================================================================

Magnetic resonance imaging (MRI) reconstruction has undergone significant transformation with the introduction of machine learning (ML) algorithms. Modern ML-based methods consistently outperform classical approaches \cite{safari_advancing_2025}, achieving robust performance even under challenging conditions including high noise level and aggressive undersampling. Despite these advances, ML models still exhibit a black-box nature \cite{chatterjee_unboxing_2023}, which makes them difficult to interpret, especially for large models with an extended number of parameters. This uncertainty creates substantial barriers to clinical translation when interacting with regulatory bodies and undermines acceptance among medical professionals \cite{dinsdale_challenges_2022}.

A clear understanding of algorithmic behavior remains a key requirement for interpretability \cite{champendal_scoping_2023}, particularly because ML-based reconstructions can introduce novel artifact types, such as blurring, excessive smoothing, and, in rare cases, hallucinations \cite{bhadra_hallucinations_2021, tivnan_hallucination_2024}. These phenomena complicate expert evaluation and radiological feedback \cite{md_hossain_systematic_2024} and may contradict established qualitative expectations developed over years of classical parallel imaging (PI) with methods such as SENSE \cite{pruessmann_sense_1999}, GRAPPA \cite{griswold_generalized_2002}, CAIPIRINHA \cite{breuer_controlled_2005}, or their state-of-the-art extensions such as Compressed Sensing \cite{lustig_sparse_2007, lustig_compressed_2008}, LORAKS \cite{haldar_low-rank_2014}, P-LORAKS \cite{haldar_p-loraks_2016}, and SPIRiT \cite{lustig_spirit_2010}.

Rigorous quantification of image quality and noise amplification is essential to address these challenges. Conventional evaluation metrics for ML reconstructions—such as normalized mean squared error (NMSE), structural similarity index (SSIM), and peak signal-to-noise ratio (PSNR) \cite{wang_image_2004}—rely on gold standard reference images, which are rarely available in routine clinical scans. These aggregate metrics, inherited from classical image processing and natural image ML, often fail to capture localized variations and instead provide a global summary of image quality. This limitation significantly impacts diagnostic utility \cite{mason_comparison_2020}, as clinical evaluations frequently focus on fine structural details rather than overall global fit. Conversely, traditional linear reconstruction techniques utilize well-established g-factor maps for SENSE \cite{pruessmann_sense_1999} and GRAPPA \cite{breuer_general_2009} to spatially characterize noise amplification. The g-factor is derived solely from calibration data, routinely acquired during clinical scans, using the linear system matrix built from kernel weights and sensitivity profiles for GRAPPA and SENSE, respectively. This fundamental methodological discrepancy further widens the gap between contemporary ML evaluation approaches and clinical feasibility.

Recently, \citet{akcakaya_scan-specific_2019} introduced a scan-specific k-space interpolation network, termed RAKI, which requires neither large annotated datasets nor fully sampled full resolution data for training. This sharply contrasts with variational networks (VNs) \cite{hammernik_learning_2018}, which learn image space filters from large, reference datasets as part of their unrolled architecture, enabling them to generalize across diverse anatomies and acquisition protocols. Variations of RAKI, including residual RAKI \cite{zhang_residual_2022} and iterative RAKI \cite{dawood_iterative_2023}, exhibit superior noise resilience in comparison to conventional linear reconstruction techniques such as GRAPPA. Nevertheless, in highly accelerated imaging scenarios, VNs may outperform k-space interpolation networks. This is attributed to their capacity to compensate for insufficient input data by upscaling the contributions of the prelearned spatial filters. This observation supports the motivation to incorporate a regularization term into RAKI, aiming to overcome current limitations by merging the most advantageous features of both approaches. 

Recent work by \citet{dawood_image_2025} has advanced the field by introducing an image-space formulation for k-space interpolation networks. This framework enabled the generalization of g-factor analysis to RAKI and provided insights into network behavior through the image-space representation of nonlinear activation filters. Complementing this, \citet{dalmaz_efficient_2025} performed noise propagation analysis for image-space ML reconstructions, including VNs, by developing a fast method to estimate the Jacobian for approximating noise variance. Together, these approaches aim to reestablish reconstruction noise characterization as a cornerstone for evaluating network performance instead of aggregate image metrics. Nonetheless, the coupled effects of denoising, blurring, and noise amplification remain insufficiently understood.

In this work, we propose an analytical framework that decomposes pixel-level noise variance of the entire reconstruction network into linear and nonlinear components: (1) the eigenpixel variance, capturing noise amplification originating from the pixel itself; and (2) the pixel contamination variance, capturing contributions from spatially separate pixels driven by nonlinear activations, which can manifest as apparent blurring. We utilize automatic differentiation to compute these variance maps efficiently at each training iteration, enabling runtime monitoring of reconstruction noise properties beyond traditional data fidelity.
By incorporating these variance maps into the training as learned regularizers within a novel g-factor-informed RAKI (GIF-RAKI) architecture, the network gains self-awareness of its noise amplification and blurring tendencies. This self-informed network can control the balance between denoising strength and excessive apparent blurring. Our experiments demonstrate that GIF-RAKI consistently outperforms conventional RAKI in noise suppression and image fidelity, supported by difference maps and aggregate quantitative metrics. Parts of this work have been presented at the Annual Meeting of the International Society for Magnetic Resonance in Medicine (ISMRM) 2024, Singapore \cite{homolya_explicit_2024}.

% ======================================================================
\section{Theory}
% ======================================================================

RAKI is briefly introduced to outline the k-space interpolation problem and establish the necessary notation. For a comprehensive discussion of RAKI in k-space and image space, the reader is referred to previous works \cite{akcakaya_scan-specific_2019, zhang_residual_2022, dawood_iterative_2023, dawood_image_2025}. We then present the formulation for variance map calculation, including its decomposition into linear and nonlinear components. Finally, we describe the integration of these variance maps into the training loss, resulting in our GIF-RAKI architecture.

\subsection{Review of k-space interpolation networks}

RAKI is a scan-specific k-space interpolation network that synthesizes missing k-space signals from acquired multichannel data. This shallow convolutional neural network (CNN) with relatively few parameters relies exclusively on enforcing k-space data consistency, effectively acting as a nonlinear generalization of GRAPPA. Unlike conventional ML methods, RAKI training is performed solely on the fully sampled low-resolution central k-space region of every slice of each scan, known as autocalibration signal (ACS). In the reconstruction phase, the network is expected to generalize to higher-frequency k-space components. The final magnitude image is then obtained by applying an inverse Fourier transform (IFFT) followed by coil channel combination.

Let $\mathbf{S}^{(0)} \in \mathbb{C}^{n_y \times n_x \times n_c^{(0)}}$ denote the acquired, undersampled, zero-filled k-space, where  $n_y, n_x,$ and $n_c^{(0)}$ represent the phase-encoding, readout, and physical input channel dimensions, respectively.
Formulated as a generalized version of GRAPPA, modern RAKI implementations use convolutional layers and nonlinear activations that operate consecutively in k-space, making the network an end-to-end k-space interpolation model. The operation of the $k^{th}$ layer is given by

\begin{equation}
    \mathbf{S}^{(k)} = \mathbb{C}\text{LReLU}^{(k)}\left(\mathbf{S}^{(k-1)}\circledast \mathbf{W}^{(k)}\right)
\end{equation}

where $\mathbf{S}^{(k)} \in \mathbb{C}^{n_y \times n_x \times n_c^{(k)}}$ is the output signal and $\mathbf{W}^{(k)} \in \mathbb{C}^{b_y^{(k)} \times b_x^{(k)} \times n_c^{(k)} \times n_c^{(k-1)}}$ is the convolution weight of the $k^{th}$ layer. Here, $\left[b_y^{(k)}, b_x^{(k)}\right]$ denote the convolution kernel sizes in phase-encoding and readout dimensions, respectively, while $\left[n_c^{(k)}, n_c^{(k-1)}\right]$ specify the filter dimensions of the $k^{th}$ and $(k-1)^{th}$ convolutional layers.
$\mathbb{C}$LReLU$^{(k)}$ abbreviates the complex-valued Leaky ReLU function \cite{cole_analysis_2021} with adjustable steepness parameter $\alpha^{(k)}$ for the $k^{th}$ layer, and $\circledast$ denotes the complex-valued convolution operation \cite{cole_analysis_2021,virtue_complex-valued_2019}. Generally, $\alpha$ determines the nonlinearity of the CNN. $\alpha = 0$ yields maximum nonlinearity as in standard RAKI, whereas $\alpha = 1$ represents a multilayer, linear GRAPPA reconstruction.

Building upon the RAKI framework, iterative RAKI (iRAKI) addresses the challenge of limited training data by augmenting the original ACS lines with synthetic ones generated via an initial GRAPPA reconstruction. This expanded training dataset enables the use of larger input kernels, facilitating improved feature extraction. The training data are iteratively refined by the previous model’s predictions. Importantly, iRAKI focuses only on refining the training data to mitigate noise amplification.

Image-space RAKI \cite{dawood_image_2025} offers a human-interpretable representation of the reconstruction process of a pretrained k-space interpolation network. Although training is fully conducted in k-space, the network's building blocks are transferred to image space for the reconstruction step, yielding quasi-equivalent image quality. By reformulating the $\mathbb{C}\text{LReLU}$ activation as a masking operation \cite{dawood_image_2025}, the reconstruction chain can be conceptualized as a cascade of convolutions and point-wise multiplications:

\begin{equation}
    \mathbf{S}^{(k)} = \mathbf{A}^{(k)}\odot\left(\mathbf{S}^{(k-1)}\circledast \mathbf{W}^{(k)}\right)
\end{equation}

where $\mathbf{A}^{(k)}\in \mathbb{C}^{n_y \times n_x \times n_c^{(k)}}$ is the activation mask of the $k^{th}$ layer, and $\odot$ denotes the complex-valued element-wise multiplication. Leveraging the convolution theorem \cite{bracewell_fourier_1986}, these operations are interchangeable between image and frequency spaces \cite{breuer_general_2009, dawood_image_2025}. This relationship equivalently translates to

\begin{align}
    \mathbf{S}^{(k)} &= \mathbf{A}^{(k)} \odot \left(\mathbf{S}^{(k-1)} \circledast \mathbf{W}^{(k)}\right) \\
    &\Updownarrow \text{ (I)FFT} \nonumber \\
    \mathbf{\hat{S}}^{(k)} &= \mathbf{\hat{A}}^{(k)} \circledast \left(\mathbf{\hat{S}}^{(k-1)} \odot \mathbf{\hat{W}}^{(k)}\right)
\end{align}

where $\mathbf{\hat{S}}^{(k)}, \mathbf{\hat{A}}^{(k)}, \mathbf{\hat{W}}^{(k)}$ represent the signal, activation mask, and convolution weights in image space for the $k^{th}$ layer, respectively.

The transition to image space enables quantification of network-induced noise enhancement without requiring direct comparison to a fully sampled gold standard \cite{dawood_image_2025}. Analogous to the g-factor in conventional parallel imaging \cite{pruessmann_sense_1999, breuer_general_2009}, the generalized RAKI g-factor $\mathbf{g} \in \mathbb{R}^{n_y \times n_x}$ \cite{dawood_image_2025} at spatial location $(i,j)$, denoted as  $g^{ij} \in\mathbb{R}$, is defined as

\begin{equation}
    g^{ij} = \frac{1}{\sqrt{R}}\frac{snr^{ij}_{\text{full}}}{snr^{ij}_{\text{acc}}}=\frac{1}{\sqrt{R}}\frac{\sqrt{\sigma^{ij}_{\text{acc}}}}{\sqrt{\sigma^{ij}_{\text{full}}}}\label{eq:gfactor_def}
\end{equation}

where $R$ denotes the equidistant undersampling factor, and $snr^{ij}_{\text{full,acc}}\in \mathbb{R}$ and $\sigma^{ij}_{\text{full,acc}}\in \mathbb{R}$ represent the voxel-wise signal-to-noise ratio (SNR) and variance of the fully sampled and accelerated images, respectively. This universal definition is subject to the reconstruction-specific variance calculation. \citet{dawood_image_2025} presented three distinct methods for computing the RAKI reconstruction-specific variance, all of which produced consistent g-factor maps. Their validation involved comparing: 1) an analytical formula, 2) numerical calculation using autodifferentiation originally introduced by \citet{wang_estimating_2022}, and 3) Monte Carlo simulations employing pseudo multiple replicas \cite{robson_comprehensive_2008}. Notably, Monte Carlo simulations served as the empirical benchmark for \citet{dalmaz_efficient_2025} as well, where no theoretical reference is available for VN noise amplification.

The voxel-wise variance $\sigma^{ij}_{\text{acc}} \in \mathbb{R}$ of the reconstructed, coil-combined, real-valued output voxel $\hat{s}_{\text{rec}}^{ij} \in \mathbb{R}$ at spatial location $(i,j)$ can be approximated to first order \cite{wang_estimating_2022} using the system Jacobian $\mathbf{J}^{ij} \in \mathbb{C}^{n_y \times n_x \times n_c^{(0)}}$ defined as:

\begin{equation}
    \mathbf{J}^{ij}=\frac{\partial\hat{s}_{\text{rec}}^{ij}}{\mathbf{\partial\hat{S}}^{(0)}}
    \label{eq:calculation_J}
\end{equation}

Following consistent notation adopted from previous works \cite{goodfellow_deep_2016}, scalars are denoted by lowercase letters, while matrices are denoted by uppercase letters. Input array indices to be summed are denoted in lowercase, while output pixel coordinates are indicated in uppercase. To facilitate compact expressions, this work employs the Einstein summation convention \cite{pfeiffer_introduction_2023}. This notational shortcut implies summation over any lowercase index in our convention that appears twice in a term, eliminating the need to write summation symbols explicitly. Accordingly, the voxel-level noise variance is expressed as:

\begin{equation}
\begin{aligned}
    \sigma^{ij}_{\mathrm{acc}} 
    = &\sum_{m=1}^{n_y}\sum_{n=1}^{n_x}\sum_{k=1}^{n_c^{(0)}}  
        \mathbf{J}^{ij}_{mnk} \boldsymbol{\Sigma}^2_{kk} 
        \left(\mathbf{J}^{ij}\right)^{\dagger}_{mnk}
     \\
    = &\sum_{m,n,k} \mathbf{J}^{ij}_{mnk} \boldsymbol{\Sigma}^2_{kk}  \left(\mathbf{J}^{ij}\right)^{\dagger}_{mnk}
     \\
    \underset{\text{Einstein}}{=} & \mathbf{J}^{ij}_{mnk} \boldsymbol{\Sigma}^2_{kk} \left(\mathbf{J}^{ij}\right)^{\dagger}_{mnk}
\end{aligned}
\label{eq:total_var}
\end{equation}

where $\dagger$ represents the Hermitian adjoint and $\boldsymbol{\Sigma}^2 \in \mathbb{C}^{n_c^{(0)} \times n_c^{(0)}}$ denotes the noise covariance matrix. Thus, the voxel-level g-factor $g^{ij} \in \mathbb{R}$ \cite{breuer_general_2009} yields

\begin{equation}
    g^{ij} 
    = \frac{1}{\sqrt{R}} \frac{\sqrt{\sigma^{ij}_{\text{acc}}}}{\sqrt{\sigma^{ij}_{\text{full}}}} 
    = \frac{1}{\sqrt{R}} \sqrt{
        \frac{
            \left(\mathbf{p}^{ij}_{k} \mathbf{J}_{mnk}^{ij}\right) \boldsymbol{\Sigma}^2_{kk} \left(\mathbf{p}^{ij}_{k} \mathbf{J}_{mnk}^{ij}\right)^{\dagger}
        }{
            \left(\mathbf{p}^{ij}\right)^T_{k} \boldsymbol{\Sigma}^2_{kk} \left(\mathbf{p}^{ij}\right)^{\ast}_{k}
        }
    }
    \label{eq:gfactor_J}
\end{equation}

where $\ast$ denotes the conjugate and $T$ the transpose operations. Computing the Jacobian after coil combination makes the coil combination weights $\mathbf{p}^{ij} \in \mathbb{C}^{n_c^{(0)}}$ redundant in the numerator, since the Jacobian $\mathbf{J}^{ij}$ already encodes this information \cite{dawood_image_2025}. The coil combination weights in the denominator $\mathbf{p}^{ij}$ are equal to one as the variance level of the fully sampled image \cite{breuer_general_2009} is already considered by the Jacobian calculation in Equation~\ref{eq:calculation_J}. The noise covariance matrix $\boldsymbol{\Sigma}^2$, however, can be omitted following noise decorrelation (i.e., noise prewhitening) of the raw k-space data \cite{pruessmann_advances_2001}. Under these conditions the voxel-level g-factor further simplifies as

\begin{equation}
  g^{ij} 
  = \frac{1}{\sqrt{R}} \frac{\sqrt{\sigma^{ij}_{\text{acc}}}}{\sqrt{\sigma^{ij}_{\text{full}}}} 
  = \frac{1}{\sqrt{R}} \,
    \sqrt{
        \mathbf{J}^{ij}_{mnk} \left(\mathbf{J}^{ij}\right)^{\dagger}_{mnk}
        }
    \label{eq:gen_gfactor}
\end{equation}

\subsection{Decomposing pixel-level noise variance}

\begin{figure}[H]
\includegraphics[width=\textwidth]{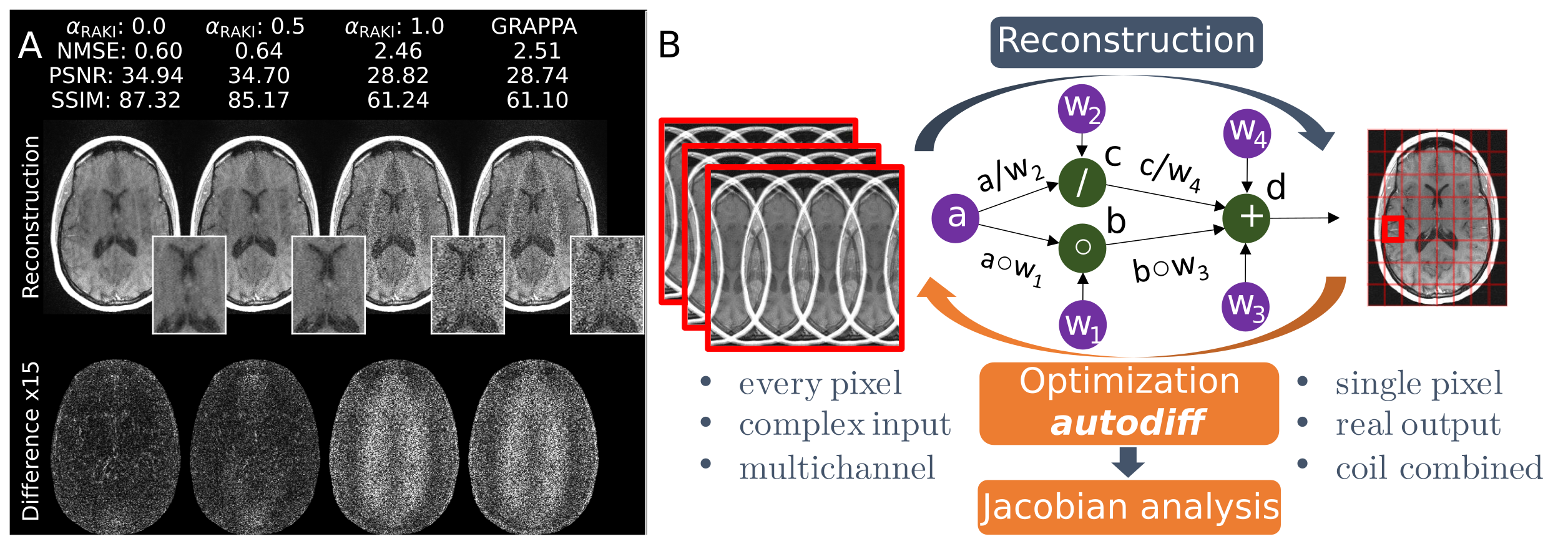}
\capt[Denoising vs. apparent blurring: a trade-off for nonlinear reconstructions.]{(A) Nonlinear RAKI reconstruction ($R=5$) with $\mathbb{C}\text{LReLU}$ factor $\alpha = 0.0$ demonstrates noticeable apparent blurring in the central region, revealing ventricular structures in the difference maps. Conversely, linear reconstructions with $\mathbb{C}\text{LReLU}$ factor $\alpha=1$ and GRAPPA show pronounced noise amplification. An intermediate nonlinearity level ($\alpha = 0.5$) achieves the best balance between effective denoising and acceptable apparent blurring. (B) Schematic representation of pixel-level variance analysis using backpropagation. The reconstruction serves as the forward model, while variance analysis exploits the backward gradient pass also used during optimization. The pixel-level variance of a single, real-valued output pixel is computed with respect to all pixels of the multichannel, complex input via built-in autodifferentiation, fully compatible with the network optimizer. This enables establishing an end-to-end connection between input and output pairs describing network noise characteristics.}  
\label{fig:Figure1_motivation}
\end{figure}

A central focus of this study is the decomposition of pixel-level noise variance into distinct linear and nonlinear components. Nonlinear activation functions effectively serve as denoising filters, but their use often introduces varying degrees of blurring or oversmoothing, as demonstrated in Figure~\ref{fig:Figure1_motivation}A with nonlinear RAKI examples for $\alpha=0$ and $0.5$. In contrast, approaches such as linear RAKI ($\alpha=1$) or GRAPPA do not exhibit visible blurring, but instead tend to enhance noise. Stronger nonlinearity appears to promote oversmoothing, whereas linear methods typically amplify Gaussian noise. To better understand and balance these competing effects—denoising versus apparent blurring—this study isolates and characterizes the contributions from linear and nonlinear signal processing within the reconstruction architecture (Figure~\ref{fig:Figure1_motivation}B) to the image of the generalized g-factor.

1) The \textit{total variance}, i.e., $\sigma^{ij}_{\text{total}} = \sigma^{ij}_{\text{acc}}$, indicates that the variance properties—and consequently the reconstruction process—of a single output pixel $\hat{s}_{\text{rec}}^{ij} \in \mathbb{R}$ depend on all input pixels $\mathbf{\hat{S}}^{(0)} \in \mathbb{C}^{n_y \times n_x \times n_c^{(0)}}$. This relationship is established through the pixel-level Jacobian $\mathbf{J}^{ij} \in \mathbb{C}^{n_y \times n_x \times n_c^{(0)}}$. The generalized g-factor $g^{ij}_{\text{total}} \in \mathbb{R}$ is a scaled version of the total variance (see Equation~\ref{eq:total_var} and \ref{eq:gen_gfactor}).

2) To elucidate the structure of the Jacobian (Equation~\ref{eq:calculation_J}), we decompose $\mathbf{J}^{ij}$ into a \textit{maximum component} and \textit{residual component}s, reflecting distinct noise intensity sources and spatial profiles. The maximum variance, $\sigma^{ij}_{\text{max}}\in \mathbb{R}$, corresponds to the largest element of $\mathbf{J}^{ij}$, identifying the input pixel $\mathbf{\hat{s}}^{(0)}_{mn}\in \mathbb{C}^{n_c^{(0)}}$ with the strongest noise contribution to output pixel $\hat{s}_{\text{rec}}^{ij}$. Notably, the maximum, i.e., primary noise source, often occurs when the input-output pixel coordinate coincide, i.e., $(m,n)=(i,j)$. The \textit{eigenpixel} term $g^{ij}_{\text{eigen}} \in \mathbb{R}$ captures the linear noise characteristics governed by the privileged interdependence of each input-output pixel pair.

\begin{equation}
  g^{ij}_{\text{eigen}}
  = \frac{1}{\sqrt{R}} \frac{\sqrt{\sigma^{ij}_{\text{max}}}}{\sqrt{\sigma^{ij}_{\text{full}}}} 
  = \frac{1}{\sqrt{R}} \,
    \sqrt{
        \mathbf{J}^{ij}_{mnk} \left(\mathbf{J}^{ij}\right)^{\dagger}_{mnk}\bigg|_{ij=mn}
    }
\end{equation}

3) The residual variance, $\sigma^{ij}_{\text{res}}\in \mathbb{R}$, accounts for the remaining noise after removing the eigenpixel contribution by setting the specific Jacobian element to zero, that is setting entries $ij=mn$ to zero: $\mathbf{J}^{ij}_{mnk}\big|_{ij=mn}=0$. Introduced by nonlinear characteristics of the network, it quantifies the aggregate influence of all other input pixels $\mathbf{\hat{s}}^{(0)}_{mn}$ on the output pixel $\hat{s}_{\text{res}}^{ij}$, where $(m,n)\neq(i,j)$. The \textit{pixel contamination} term describes signal mixing from all spatial input locations as follows:

\begin{equation}
  g^{ij}_{\text{pix}}
  = \frac{1}{\sqrt{R}} \frac{\sqrt{\sigma^{ij}_{\text{res}}}}{\sqrt{\sigma^{ij}_{\text{full}}}} 
  = \frac{1}{\sqrt{R}} \,
    \sqrt{
        \mathbf{J}^{ij}_{mnk} \left(\mathbf{J}^{ij}\right)^{\dagger}_{mnk}\bigg|_{ij\neq mn}
      }
\end{equation}

\begin{figure}[H]
\includegraphics[width=\textwidth]{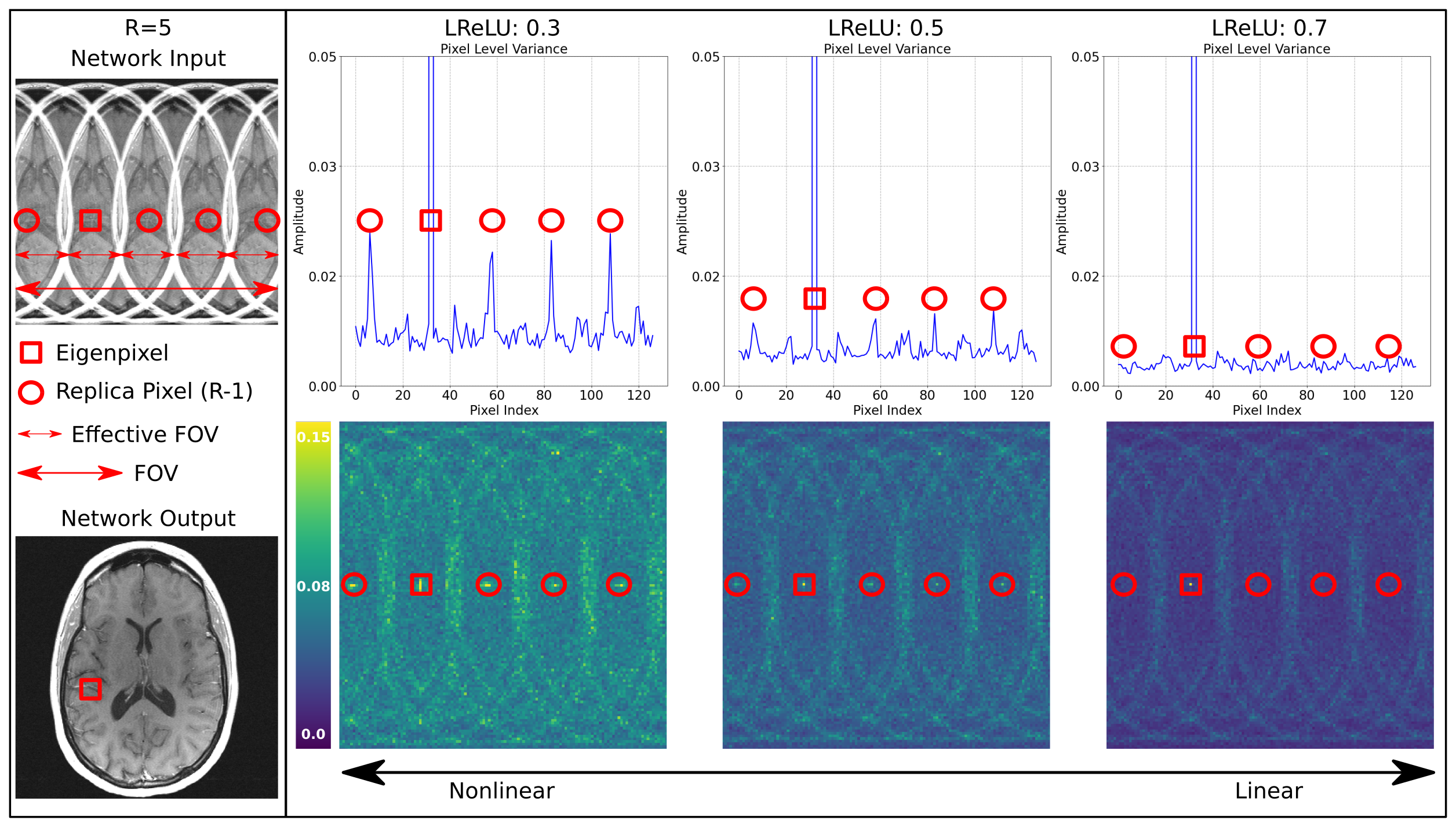}
\capt[Structure of the pixel-level variance.]{
Projected pixel-level variance maps for $\mathbb{C}\text{LReLU}$ factor $\alpha \in \{0, 0.3, 0.5, 0.7\}$ in a 5-fold acceleration setup are shown in 1D along the central line (top) and in 2D (bottom), respectively. Variance maps are reconstructed with dimensions of $127 \times 127$ after training. The examined pixel coordinate $[35;63]$, representing the eigenpixel, is outlined with a red square on all panels. Replica pixels, denoted by red circles, appear as the $(R-1)$ alias of the eigenpixel term along the PE direction, defining an effective FOV. Eigenpixel dominance is evident on the 1D plot as well as the reflected intensity pattern repeating over the effective FOV determined by the acceleration factor on the 2D images. The background magnitude increases as nonlinearity increases, enabling more signal mixing into the examined pixel. Effectively, the baseline level of the background determines the pixel contamination for the reconstruction. The widening of the eigenpixel and replica pixel peaks, which is now included in pixel contamination, is attributed to the network denoising capacity caused by introduced nonlinearity, as described by \citet{dawood_image_2025}. The nomenclature for total variance (all pixels), maximum variance (eigenpixel contribution, red squares), replica pixels (red circles), and pixel contamination (all pixels except the eigenpixel) is visually motivated.
}
\label{fig:Figure2_pixel_variance}
\end{figure}

4) The variance of the output pixel reflects the object-specific intensity pattern of the multidimensional input. Due to equidistant k-space undersampling, the $R$-fold replicated eigenpixel copies, referred to as \textit{replica pixels}, stand out in the variance landscape. Naturally, these eigenpixel replicas contribute significantly to pixel contamination. For visualization purposes (Figure~\ref{fig:Figure2_pixel_variance} and Supplementary Videos 1 and 2), the projected pixel-level variance $\boldsymbol{g}_{\text{proj}}^{ij} \in \mathbb{R}^{n_y \times n_x}$ is defined as the coil-combined pixel-level variance to demonstrate the pixel-level variance distribution similar to the spatial structure of the image-space activation mask (Figure 9 in \cite{dawood_image_2025}).

\begin{equation}
    \boldsymbol{g}_{\text{proj}}^{ij} = \frac{1}{\sqrt{R}} \,
    \sqrt{
        \mathbf{J}_{mnk}^{ij} \left( \mathbf{J}_{pqk}^{ij} \right)^{\dagger}
    }
\end{equation}

5) We define the apparent pixel blurring $g^{ij}_{\text{blur}} \in \mathbb{R}$ as the ratio of pixel contamination to the eigenpixel term, normalized by the generalized g-factor:

\begin{equation}
    g^{ij}_{\text{blur}} = \frac{g^{ij}_{\text{pix}} / g^{ij}_{\text{eigen}}}{g^{ij}_{\text{total}}}
\end{equation}

Consequently, the apparent blurring quantifies the proportion of nonlinear to linear variance, normalized by total variance, serving as a quantitative measure of visually perceived blurring. For example, $g^{ij}_{\text{blur}} = 0.5$ means the nonlinear variance is two-thirds the amplitude of the linear variance in the reconstructed pixel, whereas $g^{ij}_{\text{blur}} = 0.67$ indicates the two variance types contribute equally. For purely linear reconstructions, $g^{ij}_{\text{total}} = g^{ij}_{\text{eigen}}$ and $g^{ij}_{\text{pix}} = 0$ hold, thus $g^{ij}_{\text{blur}} = 0$ by construction. 

Extending this analysis to all pixels, the full image Jacobian $\mathbf{J} \in \mathbb{C}^{n_y \times n_x \times n_c^{(0)} \times n_y \times n_x}$ is constructed by assembling all pixel-level Jacobians $\mathbf{J}^{ij} \in \mathbb{C}^{n_y \times n_x \times n_c^{(0)}}$:

\begin{equation}
    \mathbf{J} = \nabla \mathbf{\hat{S}}^{\text{total}} = 
    \begin{bmatrix}
        \frac{\partial \hat{s}^{(1,1)}}{\partial \mathbf{\hat{S}}^{(0)}} & \dots & \frac{\partial \hat{s}^{(n_y,1)}}{\partial \mathbf{\hat{S}}^{(0)}} \\
        \vdots & \ddots & \vdots \\
        \frac{\partial \hat{s}^{(1,n_x)}}{\partial \mathbf{\hat{S}}^{(0)}} & \dots & \frac{\partial \hat{s}^{(n_y,n_x)}}{\partial \mathbf{\hat{S}}^{(0)}}
    \end{bmatrix}
\end{equation}

\begin{figure}[H]
\includegraphics[width=\textwidth]{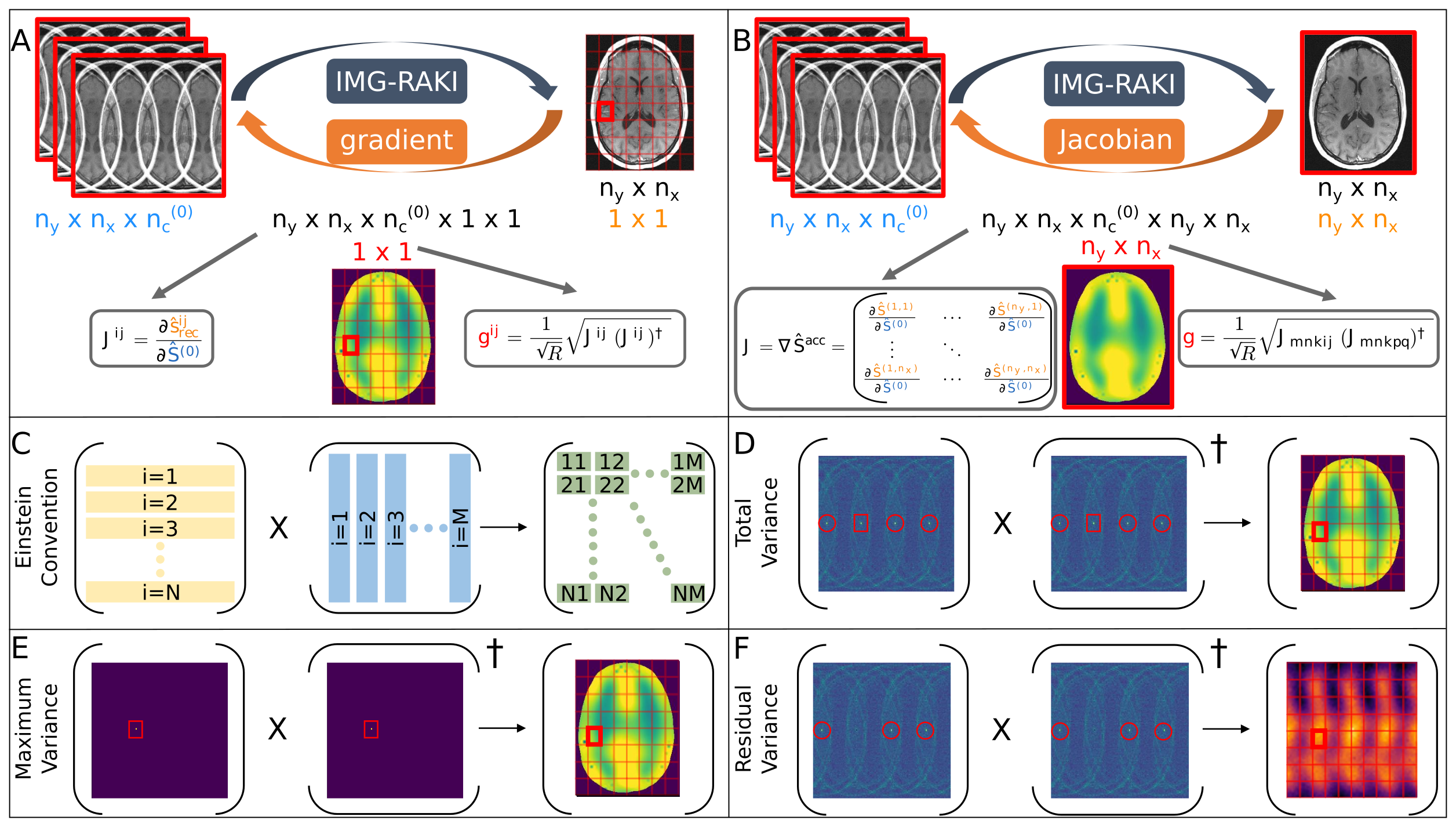}
\capt[Scalar- and vector-valued autodifferentiation and variance map nomenclature.]{
Calculating variance maps of the reconstructed output with respect to the aliased input through scalar-valued autodifferentiation (pixel gradient) using two embedded loops (A), or through vector-valued autodifferentiation (Jacobian) in a single step (B). The numerator dimensions for autodifferentiation are highlighted in orange, while the denominator dimensions are highlighted in blue for both approaches. The returned variance map component, calculated in a given step, is highlighted in red. The vector-valued calculation is considerably faster on a GPU than the scalar-valued method, at the expense of an $n_Y \times n_x$ larger memory load, as indicated by the color code. Visualization of the Einstein summation convention, omitting the channel dimension, used for fast runtime variance map calculation (C). Schematic representation of how the introduced variance maps describing the complete reconstruction output are composed from pixel-level Jacobian (D,E,F). The total variance, also known as the generalized g-factor (D), the maximum variance, termed eigenpixel (E), and residual variance, termed pixel contamination (F), map calculations are shown using the Einstein convention for 4-fold undersampling. $\dag$: Hermitian adjoint; IMG-RAKI: image-space RAKI reconstruction.
}
\label{fig:Figure3_nomenclature}
\end{figure}

This Jacobian encapsulates the end-to-end vector-valued coupling between the aliased, multi-channel, complex input and the real-valued 2D output in a single computational step (Figure~\ref{fig:Figure3_nomenclature}B). Alternatively, the gradients can be computed via pixel-wise sweeps to populate the matrix (Figure~\ref{fig:Figure3_nomenclature}A)—a low efficiency operation for GPUs. Considering all output pixels simultaneously, the pixel-level variances introduced earlier can be generalized to five dimensional tensor operations. In particular, the generalized g-factor, eigenpixel, pixel contamination, and apparent blurring maps, i.e., $\mathbf{g}_{\text{total}}, \mathbf{g}_{\text{eigen}}, \mathbf{g}_{\text{pix}}, \mathbf{g}_{\text{blur}}\in \mathbb{R}^{n_y \times n_x}$, are defined as:

\begin{align}
    \mathbf{g}_{\text{total}} &= \frac{1}{\sqrt{R}} \sqrt{\mathbf{J}_{mnkij}\left(\mathbf{J}\right)^{\dagger}_{mnkpq}} \\
    \mathbf{g}_{\text{eigen}} &= \frac{1}{\sqrt{R}} \sqrt{\mathbf{J}^{\text{max}}_{mnkij}\left(\mathbf{J}^{\text{max}}\right)^{\dagger}_{mnkpq}}\\
    \mathbf{g}_{\text{pix}} &= \frac{1}{\sqrt{R}} \sqrt{\mathbf{J}^{\text{res}}_{mnkij}\left(\mathbf{J}^{\text{res}}\right)^{\dagger}_{mnkpq}}\\
    \mathbf{g}_{\text{blur}} &= \frac{\mathbf{g}_{\text{pix}} / \mathbf{g}_{\text{eigen}}}{\mathbf{g}_{\text{total}}}
\end{align}

\begin{minipage}{0.45\textwidth}
\begin{equation}
\mathbf{J}^{\text{max}}_{mnkij} = 
\begin{cases}
\mathbf{J}_{mnkij}, & \text{if } \underset{(m,n)}{\max}|\mathbf{J}_{mnkij}| \\
0, & \text{otherwise}
\end{cases}
\end{equation}
\end{minipage}
\hfill
\begin{minipage}{0.45\textwidth}
\begin{equation}
\mathbf{J}^{\text{res}}_{mnkij} = 
\begin{cases}
0, & \text{if } \underset{(m,n)}{\max}|\mathbf{J}_{mnkij}| \\
\mathbf{J}_{mnkij}, & \text{otherwise}
\end{cases}
\end{equation}
\end{minipage}

Here, $\mathbf{J}, \mathbf{J}^{\text{max}}$, and $\mathbf{J}^{\text{res}} \in \mathbb{C}^{n_y \times n_x \times n_c^{(0)} 
\times n_y \times n_x}$ denote the total, maximum, and residual Jacobian, respectively, as illustrated in Figure~\ref{fig:Figure3_nomenclature}C, D, and F.

\subsection{G-factor informed-RAKI}

\begin{figure}[H]
\includegraphics[width=\textwidth]{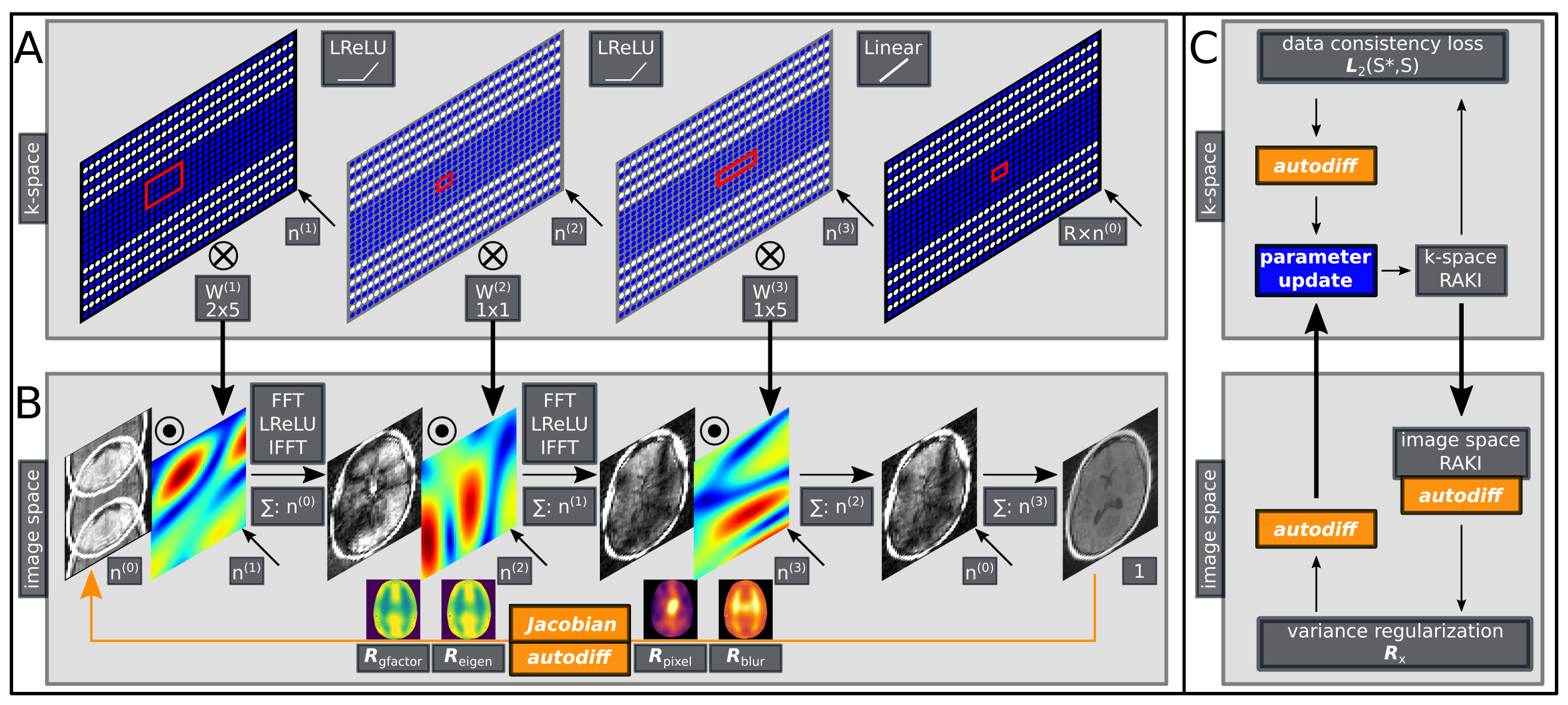}
\capt[Workflow of g-factor-informed RAKI.]{The workflow of GIF-RAKI comprises two branches: the k-space data consistency branch (A), corresponding to conventional RAKI, and an additional variance minimization term in image space (B). The schematic illustration shows network training on a fully sampled, low-resolution central k-space region (A). Low-resolution images of size $n_y^{\text{var, low}} \times n_x^{\text{var, low}} = 32 \times 32$ are derived from the undersampled k-space for the image-space branch (B). Black arrows between the two branches indicate that the updated k-space convolution weights are transferred to image space at each iteration step. Autodifferentiation is employed to compute the Jacobian and derive the variance maps based on the updated convolution weights (B). The nonlinear $\mathbb{C}\text{LReLU}$ activations are independent across branches and are retained in k-space within the image-space loop due to memory constraints. (C) A schematic overview of both branches highlights the gradient flow, illustrating the number of backpropagations through the computational graph via autodifferentiation (orange boxes). The backward pass is executed twice for the variance regularization loop (bottom, C), but only once for the k-space data consistency branch (top, C). During optimization, only the parameters of the data consistency branch are updated based on the joint k-space data consistency and image-space regularization, since the image-space branch has no additional trainable parameters of its own. Channel dimensions are omitted for clarity and indicated symbolically by arrows with dimensions.
$ \circledast$: convolution; $ \odot$: element-wise multiplication; $\Sigma : n^{(0)}$: summation in channel dimension $n^{(0)}$}
\label{fig:Figure4_workflow}
\end{figure}

GIF-RAKI is an extension of RAKI, where the network is informed in every iteration of its own noise amplification. As illustrated in Figure~\ref{fig:Figure4_workflow}, GIF-RAKI consists of two branches: 1) a k-space data consistency loop, corresponding to traditional RAKI (Figure~\ref{fig:Figure4_workflow}A), and 2) a g-factor-informed loop, where architecture-specific variances are computed in image space (Figure~\ref{fig:Figure4_workflow}B). To elucidate their distinct regularization effects, these components are incorporated separately in the total loss function $\mathcal{L}_{\text{total}}^{\scriptscriptstyle  \text{G}}\in \mathbb{R}$: 

\begin{align}
\mathcal{L}_{\text{total}}^{\scriptscriptstyle  \text{G}} &= \beta \mathcal{L}_{\text{data}}^{\scriptscriptstyle  \text{R}} + \mathcal{R}_{\text{X}} \\
\mathcal{L}_{\text{data}}^{\scriptscriptstyle  \text{R}} &= \sum |\mathbf{S}^\ast - \mathbf{S}^{(0)}_{\text{target}}|^2 \\
\mathcal{R}_{\text{X}} &= \sum \left|\mathbf{g}_{\text{X}} - \mathbf{g}_{\text{X}}^{\text{target}}\right|^2
\end{align}

Here, \(\mathcal{L}_{\text{data}}^{\scriptscriptstyle  \text{R}} \in \mathbb{R}\) denotes the RAKI data consistency loss; \(\mathcal{R}_{\text{X}}  \in \mathcal{R}_{\text{gfactor}}, \mathcal{R}_{\text{eigen}}, \mathcal{R}_{\text{pix}}, \mathcal{R}_{\text{blur}} \in \mathbb{R}\) correspond to the g-factor, eigenpixel, pixel contamination, and apparent blurring regularizers, respectively. \(\mathbf{S}^{(0)}_{\text{target}}\) and \(\mathbf{S}^\ast\) represent the target and estimated k-space data. The parameter \(\beta \in \mathbb{R}\) is an empirically set downscaling factor for the data consistency term. 

% ======================================================================
\section{Methods}
% ======================================================================

\subsection{Variance calculation using automatic differentiation}

Informing the network of its own variance properties during training necessitates a variance representation that is fast, memory-efficient, differentiable, and fully compatible with the optimizer. To accomplish this, an enhanced autodifferentiation implementation was developed in Python \cite{van_rossum_python_1995} using JAX \cite{bradbury2018compiling, bradbury_jax_2018} to tackle the higher complexity problem of the GIF-RAKI framework. JAX offers several features that make it particularly suited for our application. In PyTorch, the computational graph is built dynamically during execution by tracking operations on tensors, whereas JAX precompiles gradient functions for pure, immutable functions, enabling more efficient runtime execution. We found that JAX’s functional approach effectively mitigates vanishing gradients as opposed to PyTorch's object-oriented programming model when backpropagating several times on the computational graph. This is essential for our image-space branch employing nested autodifferentiations (Figure~\ref{fig:Figure4_workflow}C): first as a network-internal operation—similar to summation, multiplication, or convolution—used for calculating the variance maps, and second for the network parameter update. The backward pass is executed only once on the data consistency branch for parameter update. Furthermore, JAX has demonstrated superior performance over PyTorch in various benchmarks \cite{bafghi_comparing_2024, al-saeed_performance_2025}, especially when utilizing just-in-time (JIT) compilation for GPU acceleration. Importantly, in contrast to the method by \citet{dawood_image_2025}, where the Jacobian was computed after training using a stepwise, pixel-wise approach on CPU, our JAX implementation efficiently computes the Jacobian in a single step during training on GPU. While the implementation by \citet{dalmaz_efficient_2025} prioritized computational speed to derive an unbiased estimator for the Jacobian for VNs, their framework also quantified the network characteristics purely in a descriptive manner post-training using PyTorch.

\subsection{Network implementation}

All computations were performed on a single NVIDIA RTX A6000 GPU. Due to the scan-specific nature of RAKI-type reconstructions, sharing computations across GPUs along the batch dimension is typically not feasible. Furthermore, the image-space loop necessitates that the Jacobian calculation occur on the same GPU hosting the computational graph, which increases the memory requirements for GIF-RAKI relative to conventional RAKI.

All RAKI variants—RAKI, GIF-RAKI, iRAKI, and GIF-iRAKI—were implemented using complex-valued convolutions with two hidden layers comprising $n_c^{(1)}=128$ and $n_c^{(2)}=64$ channels, respectively. The output layer was assigned $n_c^{(\text{out})}=R \cdot n_c^{(0)}$, enabling the simultaneous estimation of all physical channels. The kernel sizes were $[b_y^{(1)}, b_x^{(1)}] = [2\times5]$, $[b_y^{(2)}, b_x^{(2)}]  = [1\times1]$, and $[b_y^{(\text{out})}, b_x^{(\text{out})}] =[1\times5]$. The training was performed over 500 epochs using ADAM optimizer \citep{kingma_adam_2014} with a learning rate of $5 \times 10^{-4}$. For calibration, RAKI and GIF-RAKI utilized 48 original ACS lines. iRAKI and GIF-iRAKI benefited from additional synthetic calibration lines generated via an initial GRAPPA reconstruction, resulting in 96 ACS lines with original ACS lines inserted. iRAKI and GIF-iRAKI incorporated eight iterative loops with learning rate decay of $3 \times 10^{-5}$ per iteration and employed a larger initial kernel size of $[b_y^{(1)}, b_x^{(1)}] = [4\times7]$. 

The spatial resolution of the variance maps used in GIF-RAKI was constrained by computational load and calculation time. During training, low-resolution variance maps of size $n_y^{\text{var, low}} \times n_x^{\text{var, low}}= 32 \times 32$ were employed to save resources. After training, high-resolution variance map of size $n_y^{\text{var, high}} \times n_x^{\text{var, high}} = 64 \times 64$ were computed to more accurately capture spatial variations in noise characteristics. For direct overlay visualization, these high-resolution maps were interpolated to the original image dimensions $n_y \times n_x$ using third-order spline interpolation \cite{de_boor_practical_2001}. Preliminary investigations confirmed that these resolutions are appropriate given the inherently smooth spatial nature of variance maps. Notably, the k-space inference branch did not fall under such resolution restrictions, preserving full spatial dimensions $n_y \times n_x$. To avoid additional memory allocation when convolving with image-space activation masks \cite{dawood_image_2025}, the activation functions were retained in k-space between FFT and IFFT even for the image-space branch.

We used a subset of the neuro and knee fastMRI datasets \cite{knoll_fastmri_2020} for evaluation. In the absence of available noise-only scans, k-space was prewhitened \cite{pruessmann_advances_2001} using noise covariance estimated from background region \cite{aali_gsure_2024, aali_robust_2025}. For all experiments, both reference-free quality metrics (g-factor, eigenpixel, and apparent blurring map), and gold standard-dependent aggregate metrics (NMS, SSIM, PSNR) were calculated, providing a dual perspective on image quality. The number of network parameters, floating-point operations per second (FLOPS), and runtime were also estimated.

\subsection{Observing nonlinearity through variance components}

A parameter sweep was conducted for RAKI over acceleration factors $R \in \{2, 3, 4, 5, 6\}$ and $\mathbb{C}\text{LReLU}$ parameters $\alpha \in \{0.0, 0.05, 0.1, \ldots, 1.0\}$. Variance contributions were analyzed to assess the effect of nonlinearity on reconstruction quality and its relationship with aggregate image quality metrics. Mean values of variance maps were calculated over a masked region-of-interest (ROI). We defined contour lines based on the threshold $\mathbf{g}_{\text{blur}} = 0.5$ and overlaid them on the difference and variance maps. These contour lines are intended to highlight areas where the visual perception of apparent blurring may be excessive and could potentially impact diagnostic value.

\subsection{G-factor-informed RAKI}

After initial trials of $\mathcal{R}_{\text{X}}$ regularization combinations, the joint g-factor-apparent blurring loss $\mathcal{L}_{\text{gfactor+blur}}^{\scriptscriptstyle  \text{G}}$ was developed to enable maximal denoising with minimal visually perceptible blurring defined as
\[
    \mathcal{L}_{\text{gfactor+blur}}^{\scriptscriptstyle  \text{G}} = \beta\,\mathcal{L}_{\text{data}}^{\scriptscriptstyle  \text{R}} + \mathcal{R}_{\text{gfactor}} + \mathcal{R}_{\text{blur}}
\]

$\mathcal{L}_{\text{data}}^{\scriptscriptstyle  \text{R}}$ primarily drove the optimization and was weighted with $\beta = 1 \times 10^{-4}$ to prevent it from excessively dominating the variance regularizers $\mathcal{R}_{\text{gfactor}}+\mathcal{R}_{\text{blur}}$. For the target maps, i.e., $\mathbf{g}_{\text{X}}^{\text{target}} \in \mathbf{g}_{\text{gfactor}}^{\text{target}},\mathbf{g}_{\text{blur}}^{\text{target}} \in \mathbb{R}^{n_y \times n_x}$ , we could exploit the prior knowledge of the spatial distribution of an initial GRAPPA g-factor $\mathbf{g}_{\scriptscriptstyle  \text{GRAPPA}}^{\text{target}} \in \mathbb{R}^{n_y \times n_x}$ such as 

\begin{equation}\label{eq:carcost1}
\mathbf{g}_{\text{X}}^{\text{target}} = 
\begin{cases} 
\gamma_{\text{X}} & \text{if } \mathbf{g}_{\scriptscriptstyle  \text{GRAPPA}}^{\text{target}} > \mu_{\text{X}} \\ 
0 & \text{otherwise} 
\end{cases}
\end{equation}

Here, $\mu_{\text{X}} \in \mu_{\text{gfactor}}, \mu_{\text{blur}} \in \mathbb{R}$, and $\gamma_X \in \gamma_{\text{gfactor}}, \gamma_{\text{blur}} \in \mathbb{R}$ are arbitrary parameters denoting cutoff threshold and the target value, respectively. The cutoff thresholds $\mu_{\text{gfactor}}, \mu_{\text{blur}} = 1$ were fixed for all cases, yielding homogeneous variance targets. A parameter sweep was performed with target values $\gamma_{\text{gfactor}}^{\text{target}} \in \{1.25, 1.33, 1.5, 1.67, 1.75, 2.0\}$ and $\gamma_{\text{blur}}^{\text{target}} \in \{0.33, 0.55, 0.67, 1.0\}$, exploring $R \in \{4, 5, 6\}$ and $\alpha \in \{0.3, 0.5\}$. Though, GIF-RAKI data consistency loss is identical to the RAKI one, i.e., \(\mathcal{L}_{\text{data}}^{\scriptscriptstyle \text{G}} = \mathcal{L}_{\text{data}}^{\scriptscriptstyle  \text{R}} \in \mathbb{R}\), it computes the variance metrics at every iteration step for network monitoring only. The performance of GIF-RAKI was evaluated against RAKI across various brain contrasts at group level using batches of 30 samples comprising five scans from five subjects. GIF-RAKI was configured with $\gamma_{\text{gfactor}}^{\text{target}} = 1.25$ and $\gamma_{\text{blur}}^{\text{target}} = 0.5$, $\alpha = 0.3$, $R=5$, and $\text{ACS}=48$. Statistical significance between methods was assessed using a two-tailed Student's t-test, with the significance level set at $p < 0.05$.

% ======================================================================
\section{Results}
% ======================================================================

All results presented in the main manuscript are based on a representative T1-weighted (T1w) brain dataset. Supplementary Material presents additional examples using FLAIR brain images as well as proton density (PD) knee images. Supplementary Videos 1 and 2 visualize the composition of total variance at the pixel level, providing additional insight complementing Figure 2.

\subsection{Observing nonlinearity through variance components}

\begin{figure}[H]
\centering
\includegraphics[width=\textwidth]{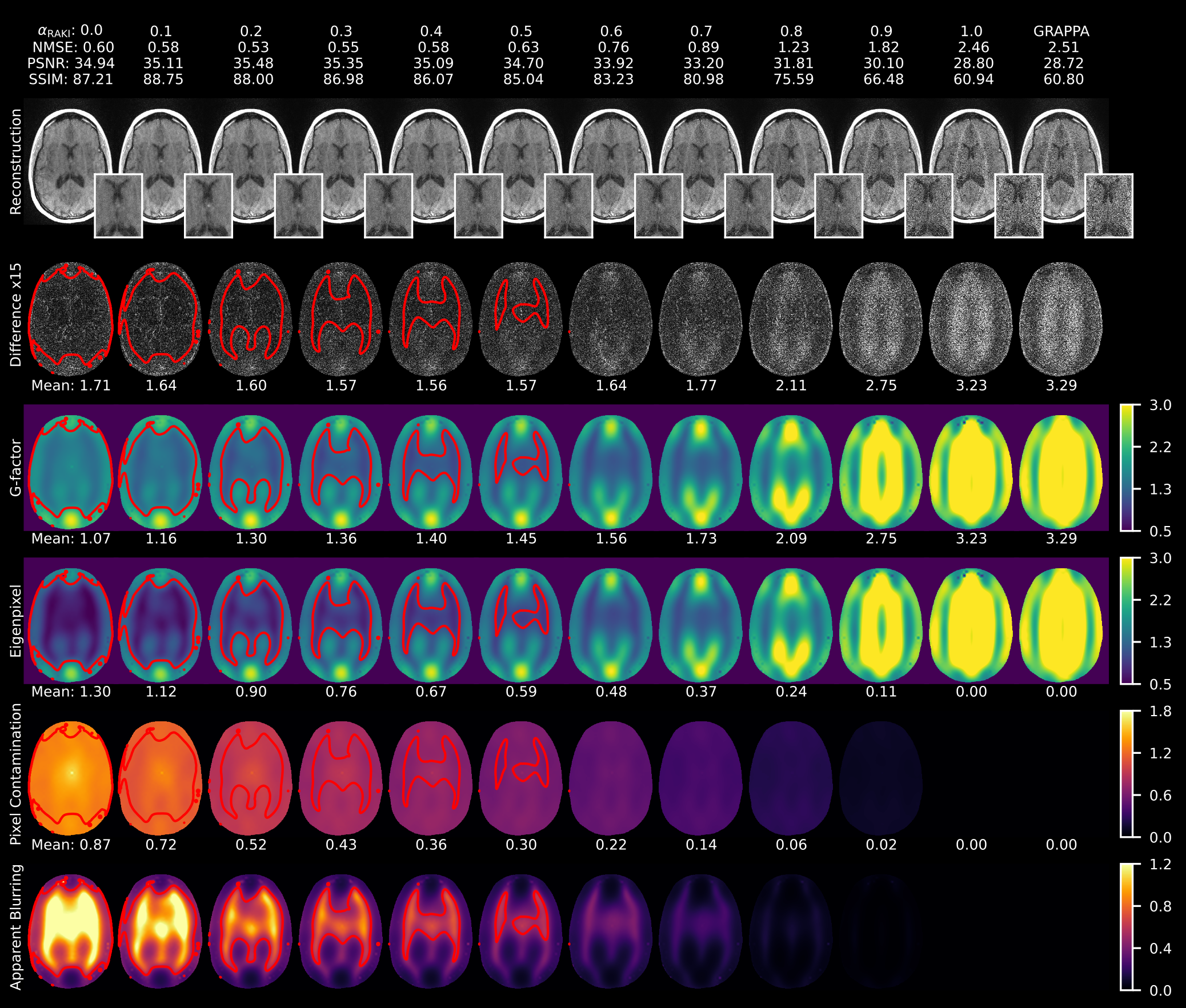}
\capt[Effect of nonlinearity on variance components.]{
RAKI and GRAPPA reconstructions, $\times 15$ magnified difference maps, generalized g-factor, eigenpixel, pixel contamination, and apparent blurring maps are shown for $\mathbb{C}\text{LReLU}$ factors $\alpha \in \{0.0, 0.1, 0.2, \ldots, 1.0\}$ with $R=5$ and $ACS=48$. Mean variances computed within the whole brain are displayed above each subfigure. The difference maps and aggregate quality metrics motivate the use of higher nonlinearity levels for improved denoising, as confirmed by the decreased g-factor values. The variance maps illustrate the composition of the g-factor: pixel contamination contributions remain negligible in the linear regime ($\alpha \approx 1.0$), but become dominant under strong nonlinearity ($\alpha \approx 0.0$), producing a noticeable blurring effect. This is reflected in the considerable drop of the eigenpixel contribution in the g-factor. An intermediate nonlinearity level ($\alpha \approx 0.5$) achieves the best balance between effective denoising and acceptable apparent blurring. The apparent blurring maps capture the visual effect of the dominant pixel contamination in the total variance, denoted by red contours for $\mathbf{g}_{\text{blur}} = 0.5$. The case $\alpha = 1$ corresponds to a RAKI network with linear activations, where pixel contamination is within numerical precision. In this purely linear case, the total variance equals the eigenpixel map, consistent with GRAPPA behavior. NMSE: normalized mean squared error ($\times 100$), PSNR: peak signal-to-noise ratio, SSIM: structural similarity index ($\times 100$).
}
\label{fig:Figure5_effect_nonlinearity}
\end{figure}

As demonstrated in Figure~\ref{fig:Figure5_effect_nonlinearity} for the nonlinearity sweep $\alpha \in \{0.0, 0.1, 0.2, \dots, 1.0\}$, pixel contamination contributions remain negligible in the linear regime but become dominant under strong nonlinearity. A balanced regime emerges near $\alpha\approx 0.5$, where eigenpixel contributions are still dominant and pixel contamination provides beneficial denoising. The distinct blurring sensation is quantified by the apparent blurring maps, becoming strong in the central ROI for highly nonlinear cases but still negligible around the edges. This coincides with the strongly denoised regions visible in the difference maps, closely following the red contour line. In highly nonlinear cases, the eigenpixel maps may locally drop below one, while an increase in pixel contamination indicates that it compensates for this suppressed eigenpixel signal contribution.

\begin{figure}[H]
\centering
\includegraphics[width=0.9\textwidth]{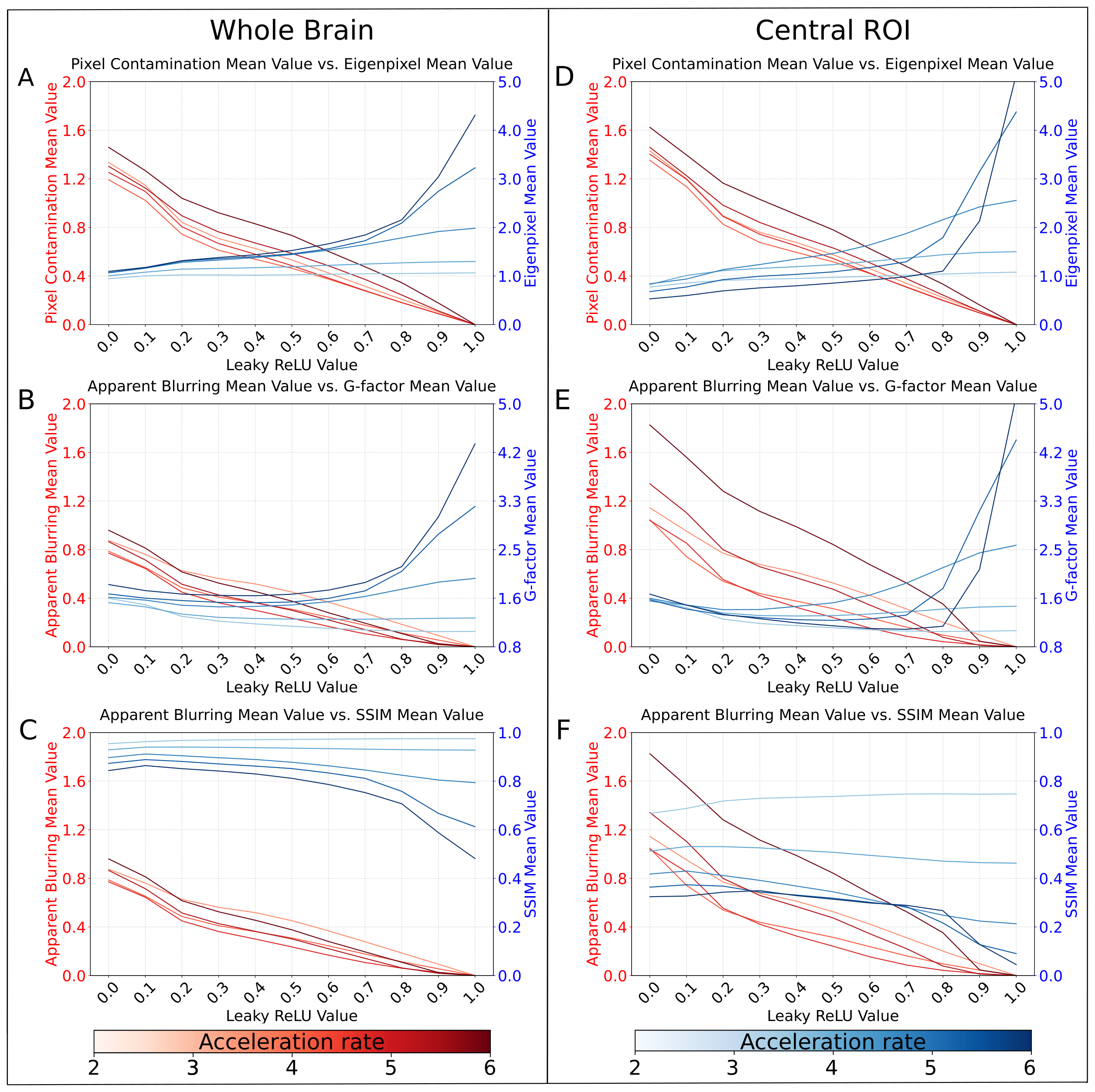}
\capt[Analysis of variance components and SSIM across nonlinearity levels and acceleration factors.]{
The plots illustrate characteristics as a function of the nonlinearity level $\alpha \in \{0.0, 0.05, 0.10, \dots, 1.0\}$ for various undersampling factors $R \in \{2, 3, 4, 5, 6\}$ with $\text{ACS} = 48$. 
Data are categorized by spatial averaging: the whole brain (left column) and a central $64 \times 64$ ROI (right column). (A, D) Linear (blue diverging curves) and nonlinear (red converging curves) variance components. The eigenpixel contribution increases steeply with $\alpha$, particularly for high undersampling ($R \in \{4, 5, 6\}$), while pixel contamination dominates in the highly nonlinear regime ($\alpha \approx 0$) and decreases monotonically. (B, E) Total variance (blue curves) formed by the interplay of components. A minimum is observed at medium nonlinearity ($\alpha \approx 0.5$) for high $R$, whereas for lower acceleration ($R \in \{2, 3\}$), the minimum is reached at $\alpha = 1$. Apparent blurring in the ROI (E) changes more steeply than pixel contamination (D), suggesting higher sensitivity to fine changes in the central region. This relation is moderate in the whole brain (B vs. A). (C, F) SSIM and apparent blurring manifest similar dependence on image smoothness. Highly accelerated cases benefit from denoising introduced by high nonlinearity, whereas lower acceleration cases suffer from excessive apparent blurring resulting in lower SSIM. Overall, the central ROI is more susceptible to nonlinearity changes than the whole brain due to localized denoising-apparent blurring behavior.}
\label{fig:Figure6_std_components_function_nonlinearity}
\end{figure}

Similar trends are observed for acceleration factors $R \in \{2,3,4,5,6\}$ when analyzing mean values of the respective variance maps over the whole brain and the central ROI (Figure~\ref{fig:Figure6_std_components_function_nonlinearity}, left and right columns, respectively). The means of the total and linear variance increase as nonlinearity decreases, showing a strong dependence on the acceleration factor. In contrast, pixel contamination is largely independent of acceleration factor and is modulated primarily by nonlinearity level (Figure~\ref{fig:Figure6_std_components_function_nonlinearity} A and D). Due to the interplay of these competing processes, a trade-off between denoising and apparent blurring emerges for a given undersampling factor, reflected by the convex shape of the g-factor curves within the moderately nonlinear regime $\alpha \approx 0.5$. Here, the mean apparent blurring remains moderate ($g_{\text{blur}} \approx 0.5$) (Figure~\ref{fig:Figure6_std_components_function_nonlinearity} B and E). SSIM promotes image smoothness driven by high nonlinearity levels, which then degrades substantially in the linear regime, especially for high undersampling factors $R \in \{4,5,6\}$ (Figure~\ref{fig:Figure6_std_components_function_nonlinearity} E and F). This is accompanied by high apparent blurring reducing the total and linear variances. However, this denoising is not spatially uniform but occurs predominantly in the central region. Steeper curves at linear and nonlinear extremes for high undersampling factors $R \in \{4,5,6\}$ indicate this when comparing whole brain and ROI indices (left and right panels).

\subsection{Impact of variance regularization}

\begin{figure}[H]
\centering
\includegraphics[width=\textwidth]{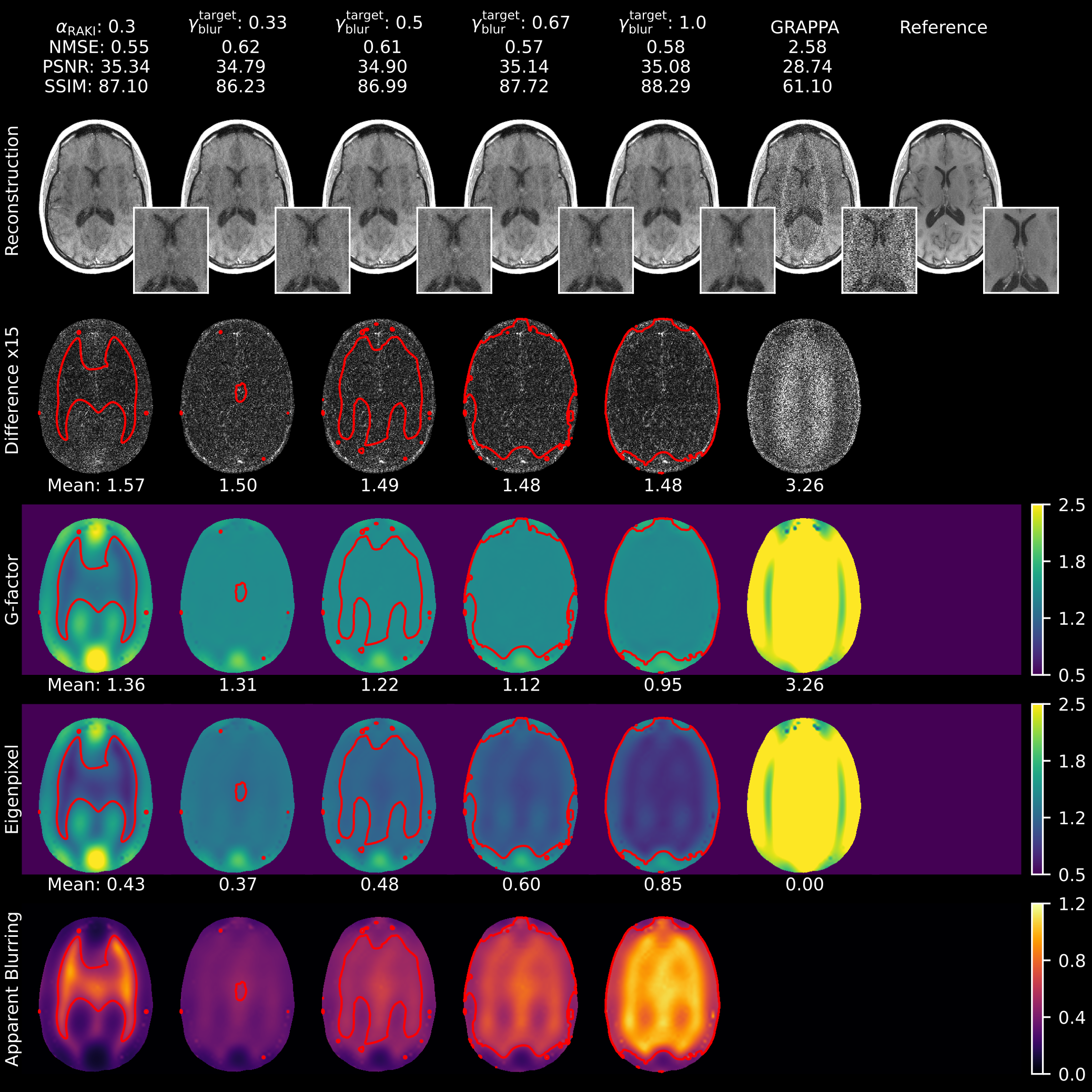}
\capt[GIF-RAKI at 5-fold undersampling: Effect of various apparent blurring levels.]{RAKI (Column 1), GIF-RAKI (Columns 2--5), and GRAPPA reconstructions alongside with reference image (Columns 6--7) are shown. Magnified difference maps ($\times 15$), g-factor, eigenpixel, and apparent blurring maps are presented for $R=5$, $\text{ACS}=48$, and $\alpha=0.3$. The g-factor target was kept constant at $\mathbf{\gamma}_{\text{gfactor}}^{\text{target}} = 1.5$, whereas the apparent blurring target was varied as $\mathbf{\gamma}_{\text{blur}}^{\text{target}} \in \{0.33, 0.5, 0.67, 1.0\}$. The visually perceived blurring on the magnitude and difference images corresponds with increasing apparent blurring levels, consistent with quality metrics. Due to variance regularization, all GIF-RAKI variance maps exhibit spatial smoothness, in contrast to RAKI, where both noisy and oversmoothed regions are to be found. Mean variances computed within the whole brain are displayed above each subfigure.
}
\label{fig:Figure7_T1w_GIF_RAKI_apparent_blurring_sweep}
\end{figure}

\begin{figure}[H]
\centering
\includegraphics[width=\textwidth]{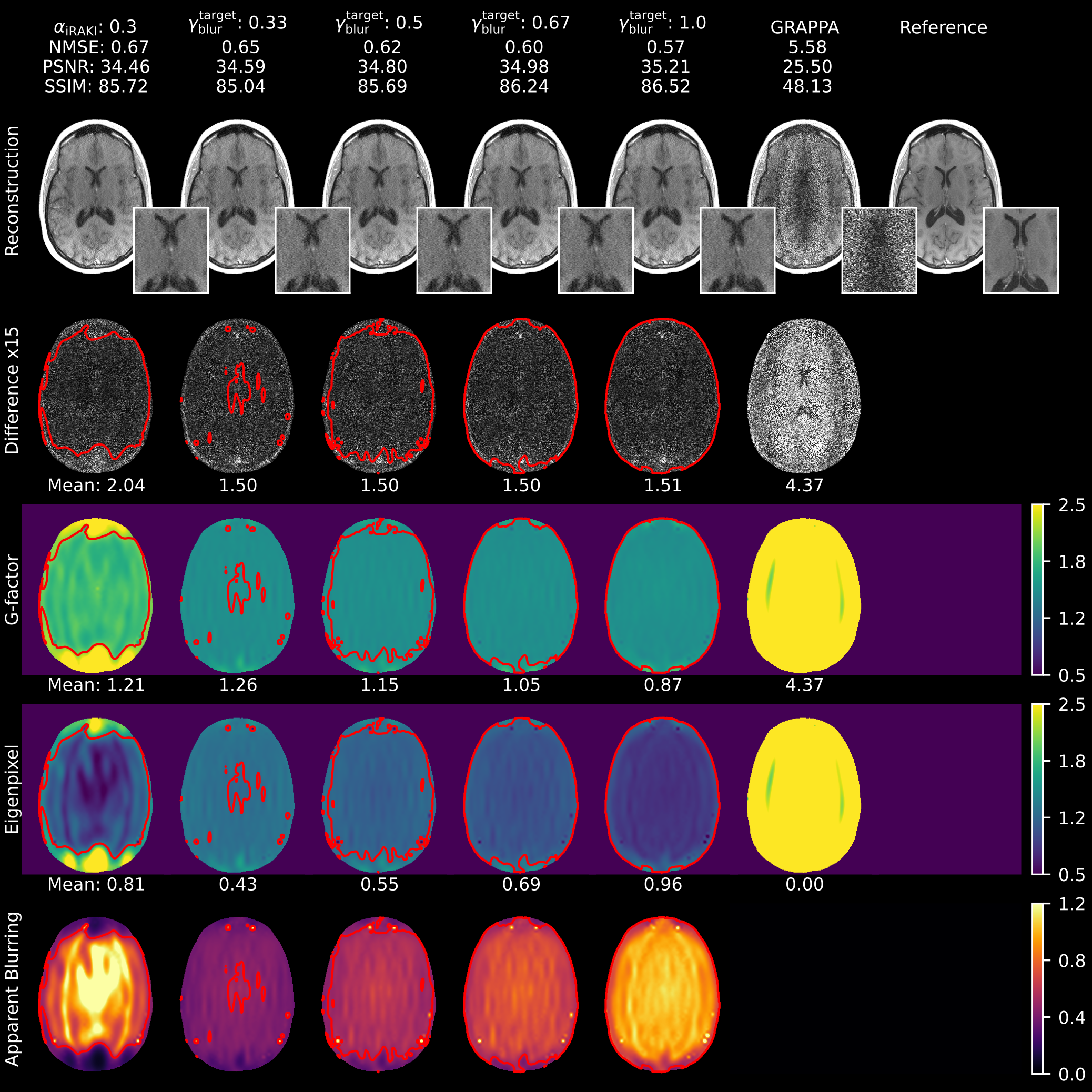}
\capt[GIF-iRAKI at 6-fold undersampling: Effect of various apparent blurring levels.]{iRAKI (Column 1), GIF-iRAKI (Columns 2--5), and GRAPPA reconstructions alongside with reference image (Columns 6--7) are shown. Magnified difference maps ($\times 15$), g-factor, eigenpixel, and apparent blurring maps are presented for $R=6$, $\text{ACS}=48$, and $\alpha=0.3$. The g-factor target was kept constant at $\mathbf{\gamma}_{\text{gfactor}}^{\text{target}} = 1.5$, whereas the apparent blurring target was varied as $\mathbf{\gamma}_{\text{blur}}^{\text{target}} \in \{0.33, 0.5, 0.67, 1.0\}$. The visually perceived blurring on the magnitude and difference images corresponds with increasing apparent blurring levels, consistent with quality metrics. Due to variance regularization, all GIF-iRAKI variance maps exhibit spatial smoothness, in contrast to iRAKI, where both noisy and oversmoothed regions are to be found. Mean variances computed within the whole brain are displayed above each subfigure.
}
\label{fig:Figure8_T1w_GIF_iRAKI_apparent_blurring_sweep}
\end{figure}

Figure~\ref{fig:Figure7_T1w_GIF_RAKI_apparent_blurring_sweep} illustrates the impact of varying apparent blurring levels at a constant g-factor target for GIF-RAKI ($R=5$). Compared to standard RAKI, GIF-RAKI variance regularization produces more homogeneous variance maps by eliminating spatially oscillating noisy and oversmoothed regions. While higher apparent blurring targets improve quantitative metrics, they also increase the perception of image smoothness. Under constant g-factor constraints, the loss function $\mathcal{L}_{\text{gfactor+blur}}^{\scriptscriptstyle \text{G}}$ effectively shifts the balance between linear and nonlinear variance components. Similar performance is observed for GIF-iRAKI at a higher 6-fold acceleration (Figure~\ref{fig:Figure8_T1w_GIF_iRAKI_apparent_blurring_sweep}). Supplementary Figures 1 and 2 demonstrate the effect of varying g-factor levels at a constant apparent blurring target for GIF-RAKI and GIF-iRAKI ($R=5$), respectively. As the g-factor rises, increased pixel contamination is required to maintain a constant level of apparent blurring. Notably, apparent blurring remains dominant even at elevated g-factor levels.

While GIF-RAKI effectively mitigates Gaussian noise, it does not account for structured artifacts arising from insufficient training data (Supplementary Figures 3 and 4). iRAKI addresses these artifacts but introduces significant noise amplification, which is mitigated by GIF-iRAKI (Supplementary Figure 5). Supplementary Figures 6 and 7 detail the influence of input layer kernel dimensions $[b_y^{(1)}, b_x^{(1)}]$ on RAKI when abundant ACS data is available ($R=6$, $\text{ACS}=96$). A $[2 \times 5]$ kernel achieves robust denoising comparable to iterative approaches (Figure~\ref{fig:Figure8_T1w_GIF_iRAKI_apparent_blurring_sweep}), while an expanded $[4 \times 7]$ kernel provides superior performance. To achieve high quality metrics, RAKI heavily relies on signal mixing manifesting as blurring hotspots.

\begin{figure}[H]
\centering
\includegraphics[width=0.9\textwidth]{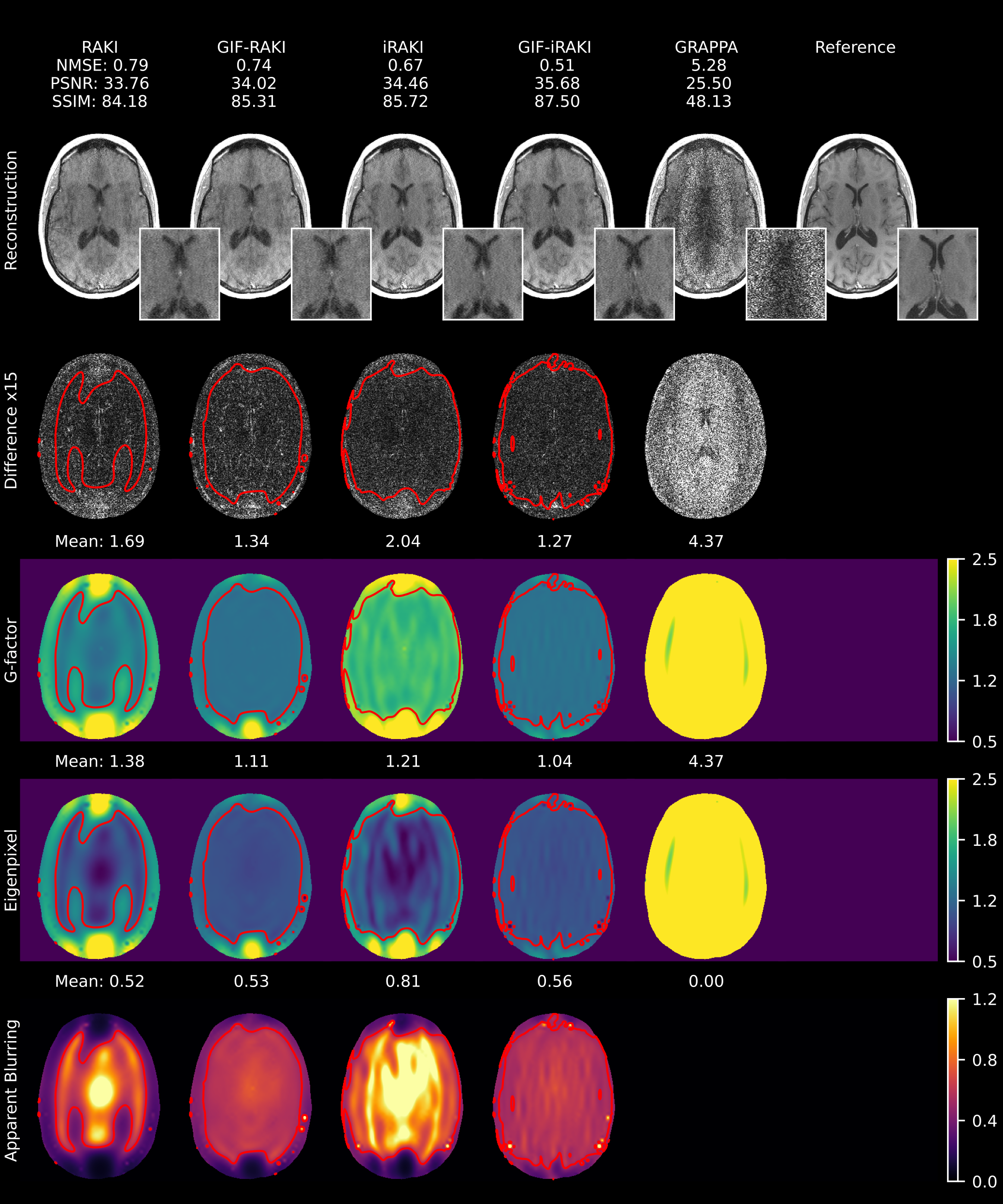}
\capt[Reconstruction performance and variance characteristics of RAKI variants.]{
RAKI, GIF-RAKI, iRAKI, GIF-iRAKI, and GRAPPA reconstructions alongside the reference image are presented. Magnified difference maps ($\times 15$), g-factor, eigenpixel, and apparent blurring maps are shown for $R=6$, $\text{ACS}=48$, and $\alpha=0.3$. For GIF-RAKI and GIF-iRAKI, the g-factor and apparent blurring targets were set to $\mathbf{\gamma}_{\mathrm{gfactor}}^{\mathrm{target}} = 1.25$ and $\mathbf{\gamma}_{\mathrm{blur}}^{\mathrm{target}} = 0.5$, respectively. Visually, apparent blurring on the magnitude and difference images is especially striking in the central hotspots for RAKI and iRAKI. The eigenpixel amplitude is suppressed, e.g., dropping below one, with foreign signal contributions compensating for it in the g-factor. Central denoised regions are bordered by noisy edges, producing an uneven variance profile. Due to spatially homogeneous variance distributions, the apparent blurring in GIF-RAKI and GIF-iRAKI is perceived more as spatially uniform denoising. Mean variances computed over the whole brain are indicated above each subfigure.
}
\label{fig:Figure9_T1w_performance_RAKI_variants}
\end{figure}

GIF-iRAKI outperforms other RAKI variants under aggressive undersampling ($R=6$, $\text{ACS}=48$), achieving a target g-factor of $\gamma_{\mathrm{gfactor}}^{\mathrm{target}} = 1.25$ as confirmed by quantitative metrics (Figure~\ref{fig:Figure9_T1w_performance_RAKI_variants}). Zoomed magnitude images reveal central blurring hotspots in standard RAKI and iRAKI that are effectively mitigated by the GIF-based variants. Furthermore, variance regularization prevents linear eigenpixel maps from falling below one. Training loss curves for GIF-iRAKI across eight refinement iterations are provided in Supplementary Figure 8.

The reconstruction times for the T1w dataset were, on average, $3.42\,\mathrm{s}$, $42.2\,\mathrm{s}$, and $91.1\,\mathrm{s}$ for $\mathcal{L}_{\text{data}}^{\scriptscriptstyle \text{R}}$, $\mathcal{L}_{\text{data}}^{\scriptscriptstyle \text{G}}$, and $\mathcal{L}_{\text{gfactor+blur}}^{\scriptscriptstyle \text{G}}$, respectively.
Supplementary Table 1 shows that the number of model parameters remains unchanged across loss functions. However, monitoring variance maps during training brings on a substantial computational cost: $\mathcal{L}_{\text{data}}^{\scriptscriptstyle \text{G}}$ requires $\approx 600\times$ more operations than $\mathcal{L}_{\text{data}}^{\scriptscriptstyle  \text{R}}$. Incorporating variance maps into the optimization, $\mathcal{L}_{\text{gfactor+blur}}^{\scriptscriptstyle \text{G}}$ adds a total of $\approx 1300\times$ increase in computational steps compared to $\mathcal{L}_{\text{data}}^{\scriptscriptstyle  \text{R}}$.

\begin{figure}[H]
\centering
\includegraphics[width=\textwidth]{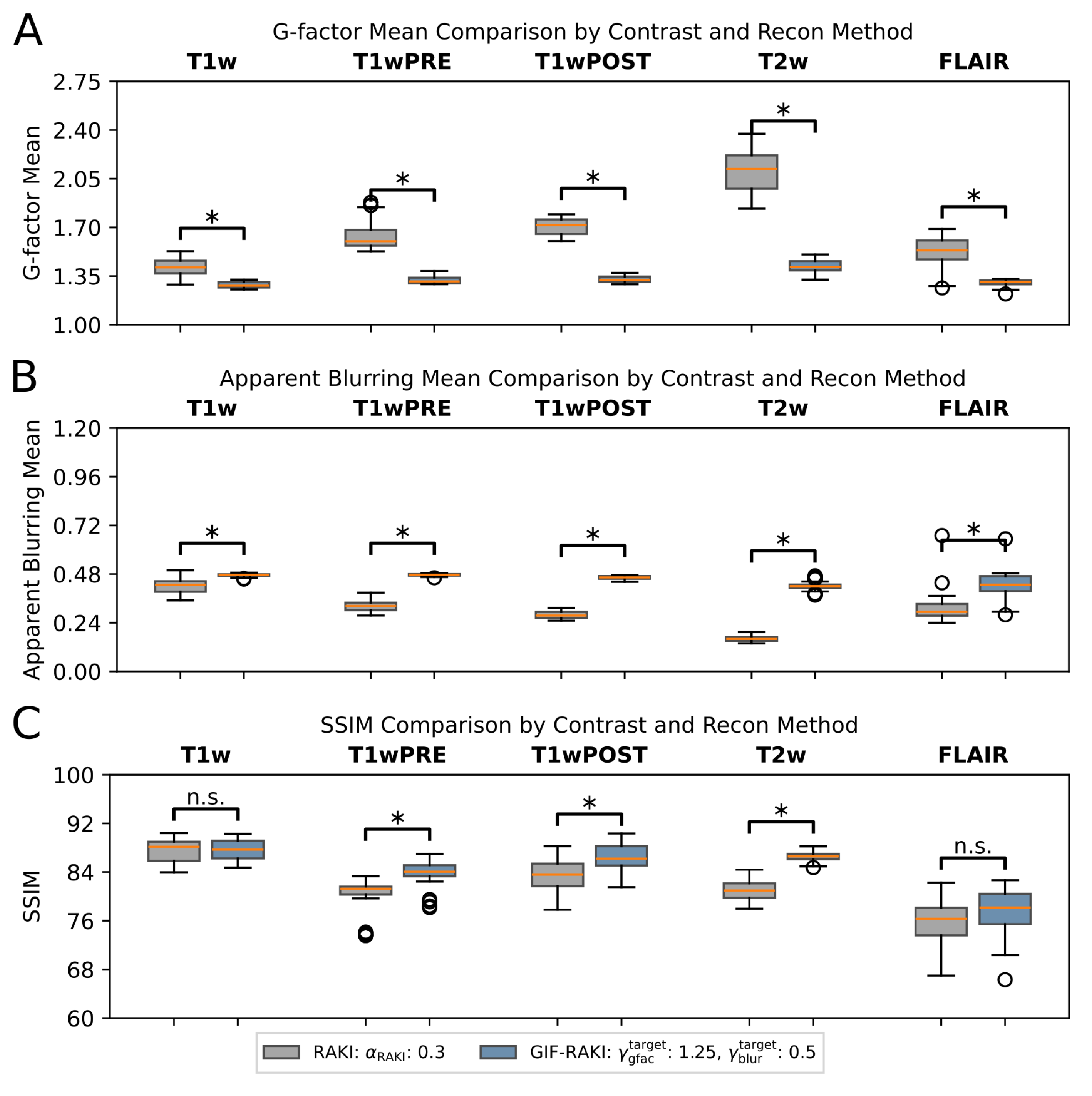}
\capt[Effect of variance regularization on image quality metrics across contrasts at the batch level.]{
Mean g-factor (A), apparent blurring (B), and SSIM (C) across brain scan contrasts at the group level. Each batch comprises 5 scans from 5 subjects yielding 30 samples in total. Compared methods include: (1) RAKI with $\alpha = 0.3, R=5, \text{ACS}=48$ and (2) identical GIF-RAKI with targets $\gamma_{\text{gfactor}}^{\text{target}} = 1.25$ and $\gamma_{\text{blur}}^{\text{target}} = 0.5$. With the given regularization targets, GIF-RAKI exhibits significantly lower mean g-factors but significantly higher apparent blurring compared to RAKI, indicating lower total noise amplification but increased signal mixing from neighboring pixels. At the batch level, SSIM follows the apparent blurring trend across all contrasts, suggesting that SSIM promotes image smoothness. Nevertheless, significant SSIM increases over RAKI are limited to T1w$_{\text{pre}}$, T1w$_{\text{post}}$, and T2w contrasts. n.s.: non-significant, $\ast$: significant with significance level set at $p < 0.05$. 
}
\label{fig:Figure10_metrics_batch}
\end{figure}

Quantitatively, GIF-RAKI demonstrated a significantly lower mean g-factor compared to RAKI accompanied by significantly higher apparent blurring at group level (Figure~\ref{fig:Figure10_metrics_batch}). SSIM values closely followed the apparent blurring trend regardless of the contrast. Statistically significant improvements in SSIM were observed in the T1w$_{\text{pre}}$, T1w$_{\text{post}}$, and T2w contrasts.

% ======================================================================
\section{Discussion}
% ======================================================================

In this work, we decomposed the pixel-level noise variance of a k-space interpolation network into linear and nonlinear contributions: (1) the eigenpixel variance, which captures noise amplification originating from the pixel location itself; and (2) the pixel contamination variance, which captures contributions from spatially separate pixel locations driven by nonlinear activations. We extended beyond intuitively motivating network denoising by quantifying the apparent blurring of the network as the normalized ratio of pixel contamination to the eigenpixel variance. Relying solely on network parameters, the variance metrics were suitable as recurrent information in a self-informed network to achieve a trade-off between denoising and apparent blurring.

The eigenpixel term exhibits a clear dependence on the acceleration factor, consistent with observations in GRAPPA, where higher acceleration leads to increased noise amplification (Figure~\ref{fig:Figure6_std_components_function_nonlinearity}A and D). Pixel contamination, on the other hand, is predominantly driven by the degree of nonlinearity and is strongly linked to the network's denoising capability (Figure~\ref{fig:Figure6_std_components_function_nonlinearity}B and E). The acceleration factor plays a secondary role in pixel contamination by effectively reducing the field-of-view through undersampling, thereby shortening the distance between replica pixels and their origin. This process facilitates the network’s increased inclusion of neighboring pixels.

Although noise amplification is significantly reduced with stronger nonlinearity, this improvement comes at the expense of increased signal contributions from foreign pixels, manifested as reduced contrast, oversmoothing, and greater apparent blurring, particularly in central image regions (Figure~\ref{fig:Figure5_effect_nonlinearity}.) As the required nonlinearity for adequate denoising increases with higher acceleration, the network increasingly relies on foreign pixels to reconstruct a single output pixel, thereby suppressing the eigenpixel term. In extremely nonlinear scenarios, the eigenpixel can drop below one, which is a non-physical scenario, suggesting that the noise level of the output pixel is lower than its input counterpart. In contrast, highly linear RAKI variants closely resemble GRAPPA in their reconstruction behavior. In these cases, image quality is primarily affected by Gaussian noise with minimal pixel contamination.

We also demonstrated that a purely data consistency-driven k-space interpolation network can be further extended with image-space variance regularizers. In this self-informed framework, the network is exposed to recurrent information regarding its own behavior. A trade-off between denoising and apparent blurring can be effectively achieved through supervised optimization of the generalized g-factor and apparent blurring maps. Due to the absence of a well-defined target for the pixel contamination maps, we chose to impose constraints not directly on the nonlinearity itself, but rather on its visual perception. Thus, apparent blurring regularization is primarily tailored for human observers rather than for setting an explicit level of nonlinearity.

Variance regularization beneficially influences image quality. Minimizing the g-factor imposes an absolute constraint on the overall noise level, while the apparent blurring regularization balances the ratio between linear and nonlinear components within the total variance (Figure~\ref{fig:Figure7_T1w_GIF_RAKI_apparent_blurring_sweep} and~\ref{fig:Figure8_T1w_GIF_iRAKI_apparent_blurring_sweep}, Supplementary Figure 1 and 2). These observations justify a regularization strategy that prioritizes achieving specific apparent blurring targets at lower g-factor levels. For the joint g-factor-apparent blurring loss $\mathcal{L}_{\text{gfactor+blur}}^{\scriptscriptstyle \text{G}}$ specifically, targets of $\gamma_{\mathrm{gfactor}}^{\mathrm{target}} = 1.25$ and $\gamma_{\mathrm{blur}}^{\mathrm{target}} = 0.5$ were proved to be effective for conservative variance regularization at the group level as well (Figure~\ref{fig:Figure10_metrics_batch}). The higher apparent blurring level imposed by variance regularization translated into elevated SSIM scores compared to RAKI. This indicates the metric's inherent bias toward image smoothness, as previously documented \cite{mason_comparison_2020, pambrun_limitations_2015}.
Furthermore, $\mathcal{L}_{\text{gfactor+blur}}^{\scriptscriptstyle \text{G}}$ encourages homogeneous variance maps and prevents the eigenpixel variance from dropping below one, thus avoiding non-physical scenarios. Simultaneously, it suppresses the network's tendency to replace noisy regions with nonlinear signal mixing from other pixel locations, thereby reducing the central blurring hotspots (Figure~\ref{fig:Figure9_T1w_performance_RAKI_variants}). 

Nevertheless, variance regularization does not eliminate the need for sufficient training data. Although GIF-RAKI effectively mitigates Gaussian noise propagation, it does not eliminate structured artifacts, such as anatomical contour lines, remaining aliasing, or the central RAKI artifact. To overcome the problem of limited training data, we combined GIF-RAKI with iterative methods to synthesize abundant training data with controlled variance levels (GIF-iRAKI).

Contrary to conventional ML approaches, we avoid expanding the hyperparameter space: GIF-RAKI requires no more parameters than standard RAKI (Supplementary Table 1). Instead, we choose to perform more computation on the image-space representation of the same model. Nevertheless, backpropagating through the computational graph twice increased the computational load and time, as well as limiting the resolution of the variance maps for the image-space branch. This approach has two major limitations: the risk of vanishing gradients due to multiple autodifferentiations and the additional demand on GPU memory.

Our goal was twofold: first, to explain the behavior of k-space interpolation networks in medical image reconstruction; and second, to incorporate this knowledge into the training process. Explainable AI approaches in the natural image ML community have also primarily relied on gradient-based methods. Grad-CAM \cite{selvaraju_grad-cam_2017} and Grad-CAM++ \cite{chattopadhay_grad-cam_2018} reveal important regions in segmentation tasks; however, these methods include only the output layer in autodifferentiation, whereas our approach monitors the entire network, enabled by the low parameter count. \citet{chatterjee_unboxing_2023} proposed teacher/parent training based on Grad-CAM segmentation boxes, feeding information from one network to another. Here, we opted for a single self-informed architecture instead.

Post-reconstruction image denoising has also been applied in medical imaging using standard denoisers. Techniques that employ scan-specific noise priors, using the GRAPPA g-factor \cite{pfaff_training_2022}, initial RAKI reconstruction \cite{mei_raki_2024}, or g-factor maps \cite{yoon_g-factor_2019}, outperform off-the-shelf denoisers, but commonly assume the noise prior is constant during reconstruction. In contrast, we provide dynamic variance information in every iteration step. In addition, GIF-RAKI has the ability to produce spatially homogeneous variance maps, possibly making off-the-shelf denoisers more effective. By eliminating the substantial variability of noisy and blurred regions, the resulting image is visually more appealing and can tolerate a higher level of apparent blurring. 

Our methodology advances both the practical and theoretical aspects of ML-based MRI reconstruction by enhancing the interpretability of network functionality while simultaneously making use of this insight during reconstruction. Including parameter minimization as a regularizer beyond the data consistency term is a common method in VNs, but the current GIF-RAKI implementation explicitly minimizes a physical quantity, the network's noise variance. Rendering a verdict on the optimal variance levels is beyond the scope of this study, as this would require systematic expert rating and radiologist feedback to achieve a robust consensus.

% ======================================================================
\section{Conclusion}
% ======================================================================

The pixel-level noise variance of a k-space interpolation network is analyzed to improve the interpretability of ML-based medical image reconstruction. The generalized g-factor is decomposed into eigenpixel and pixel contamination components, capturing the network's linear and nonlinear noise propagation, respectively. Swift calculation of the Jacobian through autodifferentiation enables runtime monitoring of network performance and integrates seamlessly with the optimizer. Extending k-space data consistency RAKI with a term that accounts for noise amplification and apparent blurring provides inherent, physics-driven regularization for the network. As a result, the self-informed GIF-RAKI network can achieve a trade-off between denoising and apparent blurring as part of the reconstruction process, even under aggressive undersampling and limited training data.

%TC:ignore
% ======================================================================
\section{Acknowledgments}
% ======================================================================

I.H. gratefully acknowledges funding by the Center for Clinical Research (IZKF) of the Medical Faculty Würzburg Project F-461 (Hein/Gamer/Terekhov), and by the Tempus Public Foundation Project MÁEÖ 2022-23/175912.
\newline
F.K. gratefully acknowledges funding by the German Research Foundation (DFG) (projects 513220538, 512819079; and project 500888779 of the RU5534 MR biosignatures at UHF).
\newline
Special thanks to Peter Kemenczky for the fruitful discussions at the early stages of this work.

% ======================================================================
\section{Conflict of Interest}
% ======================================================================

F.K. receives patent royalties for deep learning image reconstruction and research support from Siemens Healthineers AG, has stock options from Subtle Medical, and is a consultant for Imaginostics.

% ======================================================================
\section{Data Availability Statement}
% ======================================================================

In the spirit of reproducible research, the source code of GIF-RAKI is available on GitHub: \url{https://github.com/ihomolya/GIF_RAKI} 

% ======================================================================
\section{ORCID}
% ======================================================================

\textit{Istvan Homolya\hspace{2mm}}\href{https://orcid.org/0000-0002-0662-6464}{\orc}\hspace{2mm}\url{https://orcid.org/0000-0002-0662-6464} \newline
\textit{Jannik Stebani\hspace{2mm}}\href{https://orcid.org/0009-0004-9631-9928}{\orc}\hspace{2mm}\url{https://orcid.org/0009-0004-9631-9928} \newline
\textit{Grit Hein\hspace{2mm}\href{https://orcid.org/0000-0001-5696-6486}{\orc}}\hspace{2mm}\url{https://orcid.org/0000-0001-5696-6486} \newline
\textit{Matthias Gamer\hspace{2mm}}\href{https://orcid.org/0000-0002-9676-9038}{\orc}\hspace{2mm}\url{https://orcid.org/0000-0002-9676-9038} \newline
\textit{Florian Knoll\hspace{2mm}}\href{https://orcid.org/0000-0001-5357-8656}{\orc}\hspace{2mm}\url{https://orcid.org/0000-0001-5357-8656} \newline
\textit{Martin Blaimer\hspace{2mm}}\href{https://orcid.org/0000-0002-6360-9871}{\orc}\hspace{2mm}\url{https://orcid.org/0000-0002-6360-9871}

%TC:endignore
%TC:ignore
% ======================================================================
% \section{References}
% ======================================================================
%\bibliography{GIF_RAKI_References} 

%TC:endignore
%TC:ignore
% ======================================================================
\section{Figures and Tables}
% ======================================================================

%TC:endignore

\clearpage
\includepdf[pages=-]{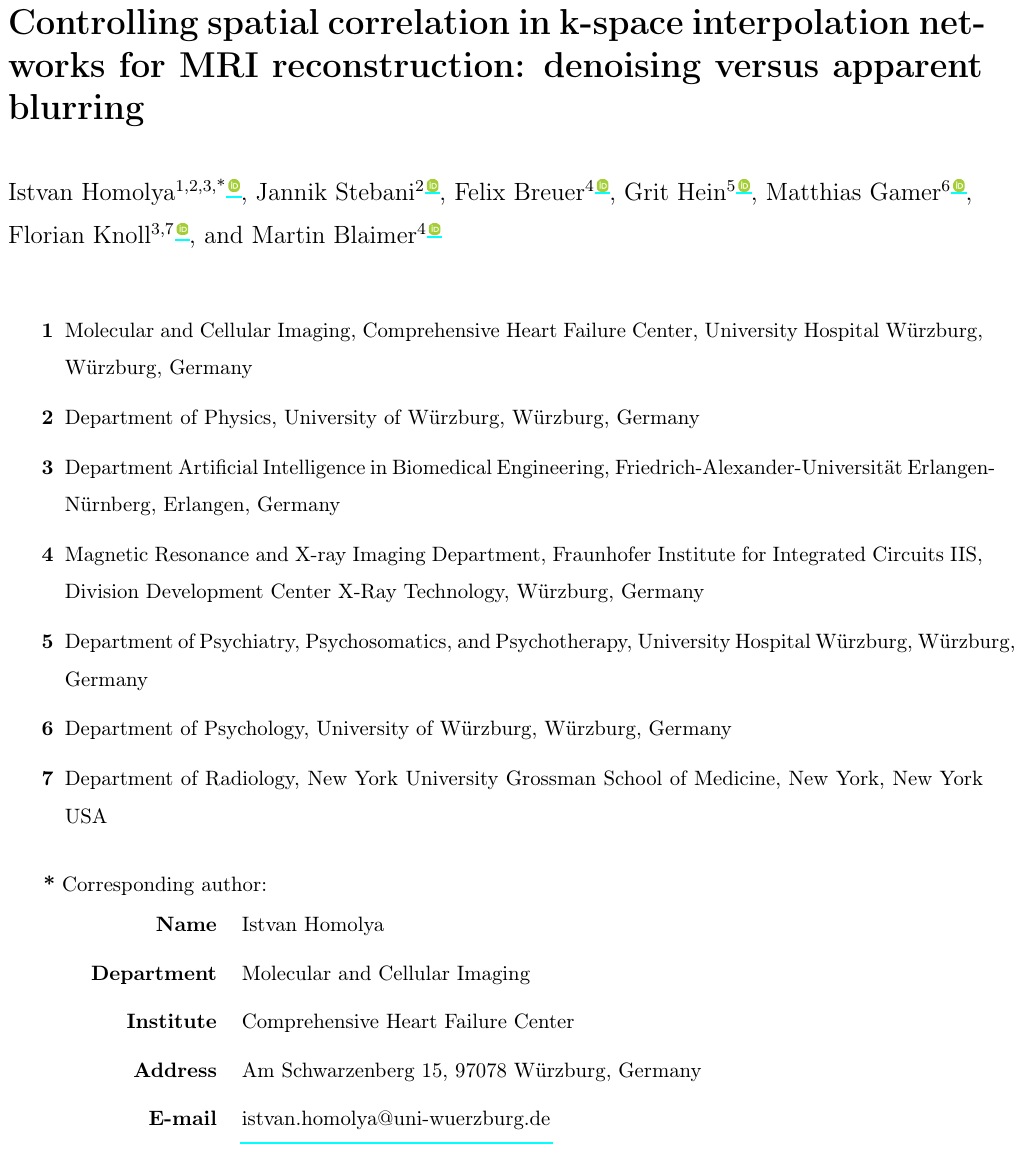}

\end{document}